\documentclass[sn-mathphys-num]{sn-jnl}% Math and Physical Sciences Numbered Reference Style 
\usepackage{graphicx}%
\usepackage{multirow}%
\usepackage{amsmath,amssymb,amsfonts}%
\usepackage{amsthm}%
\usepackage{mathrsfs}%
\usepackage[title]{appendix}%
\usepackage{xcolor}%
\usepackage{textcomp}%
\usepackage{manyfoot}%
\usepackage{booktabs}%
\usepackage{algorithm}%
\usepackage{algorithmicx}%
\usepackage{algpseudocode}%
\usepackage{listings}%
\usepackage{siunitx}%
\usepackage{subfigure,dcolumn}%

\begin{document}

\title[muEDM Muon Trigger Detector]{Beam test performance of a prototype muon trigger detector for the PSI muEDM experiment}


\author[1,2]{Tianqi Hu}
\equalcont{These authors contributed equally to this work.}

\author[1,2]{Jun Kai Ng}
\equalcont{These authors contributed equally to this work.}

\author[1,2]{Guan Ming Wong}
\equalcont{These authors contributed equally to this work.}

\author[1,2]{Cheng Chen}
\author*[1,2]{\fnm{Kim Siang} \sur{Khaw}}\email{kimsiang84@sjtu.edu.cn}
\author[1,3]{Meng Lyu}
\author[4,5,6]{Angela Papa}
\author[6]{Philipp Schmidt-Wellenburg}
\author[6]{David Staeger}
\author[7,5]{Bastiano Vitali}

\affil[1]{\orgdiv{Tsung-Dao Lee Institute}, \orgname{Shanghai Jiao Tong University}, \orgaddress{\street{No.1 Lisuo Road}, \city{Shanghai}, \postcode{201210}, \country{China}}}

\affil[2]{\orgdiv{School of Physics and Astronomy}, \orgname{Shanghai Jiao Tong University}, \orgaddress{\street{No.800 Dongchuan Road}, \city{Shanghai}, \postcode{200240}, \country{China}}}

\affil[3]{\orgdiv{School of Electronics, Information and Electrical Engineering}, \orgname{Shanghai Jiao Tong University}, \orgaddress{\street{No.800 Dongchuan Road}, \city{Shanghai}, \postcode{200240}, \country{China}}}

\affil[4]{\orgdiv{Dipartimento di Fisica}, \orgname{Università di Pisa}, \orgaddress{\street{Via B.  Pontecorvo 3}, \city{Pisa}, \postcode{56127}, \country{Italy}}}

\affil[5]{\orgdiv{Sezione di Pisa}, \orgname{Instituto Nazionale di Fisica Nucleare}, \orgaddress{\street{Largo B.  Pontecorvo 3}, \city{Pisa}, \postcode{56127}, \country{Italy}}}

\affil[6]{\orgdiv{Laboratory for Particle Physics}, \orgname{Paul Scherrer Institut}, \orgaddress{\street{Forschungsstrasse 111}, \city{Villigen PSI}, \postcode{5232}, \country{Switzerland}}}

\affil[7]{\orgdiv{Dipartimento di Fisica}, \orgname{Università di Roma ``La Sapienza''}, \orgaddress{\street{Piazzale Aldo Moro 2}, \city{Roma}, \postcode{00815}, \country{Italy}}}

\abstract{We present the design, construction, and beam test results of a prototype muon trigger detector developed for the muon electric dipole moment (muEDM) experiment at the Paul Scherrer Institute (PSI). The muEDM experiment aims to increase the sensitivity of the muon EDM measurement by several orders of magnitude beyond the current limit established by the BNL E821 experiment. Precise and reliable muon identification at the entrance of the storage solenoid is crucial, as the trigger detector must quickly generate a trigger signal to activate a pulsed magnetic kicker, enabling the capture and storage of incoming muons. The trigger detector consists of two primary components: a thin gate detector made from a plastic scintillator read out by eight silicon photomultipliers (SiPMs), and a telescope detector made of four plastic scintillators also read out by SiPMs. The telescope detector operates in anticoincidence with the gate detector, identifying muons that pass through the gate detector without activating the telescope detector, thus ensuring the selection of muons with trajectories optimal for stable storage within the solenoid. A proof-of-principle test was performed at the PSI $\pi$E1 beamline using \SI{27.5}{MeV/c} muons to characterize the detector’s timing performance, scintillator light yield, and triggering efficiency. Experimental data showed excellent agreement with detailed Geant4 Monte Carlo simulations that incorporated optical photon generation, transportation, and detection. The results successfully validate the detector design and confirm its suitability for the stringent timing, efficiency, and trajectory-selection requirements essential to the muEDM experiment.}




\maketitle

\section{Introduction}\label{sec1}

The search for the muon electric dipole moment (EDM) represents one of the most promising avenues for probing physics beyond the Standard Model (SM). The current experimental limit on the muon EDM, set by the BNL E821 experiment~\cite{Muong-2:2008ebm}, $d_{\mu} < 1.8 \times 10^{-19}$~$e\cdot\text{cm}$ (95\% confidence level), is several orders of magnitude higher than the SM's prediction~\cite{Pospelov:2013sca, Ghosh:2017uqq, Yamaguchi:2020eub}. By reaching an unprecedented sensitivity of $6 \times 10^{-23}$~$e\cdot\text{cm}$, the PSI muEDM experiment~\cite{Adelmann:2025nev, Cavoto:2023xtw, PSImuEDM:2023dsd} has the potential to reveal new physics, including hints of undiscovered particles or forces.

The significance of the lepton EDM searches stems from its potential to address key unresolved questions in modern physics, such as the baryon asymmetry of the universe~\cite{Sakharov:1967dj, WMAP:2010qai} via electroweak baryogenesis~\cite{Morrissey:2012db}. Despite the remarkable success of the SM in describing particle interactions, it fails to account for the observed matter-antimatter asymmetry, suggesting the existence of additional CP-violating sources beyond the SM. Detecting a non-zero EDM would provide direct evidence of such sources and offer insight into the mechanisms underlying this asymmetry.

\begin{figure}[htbp]
\centering 
\includegraphics[width=.99\hsize]{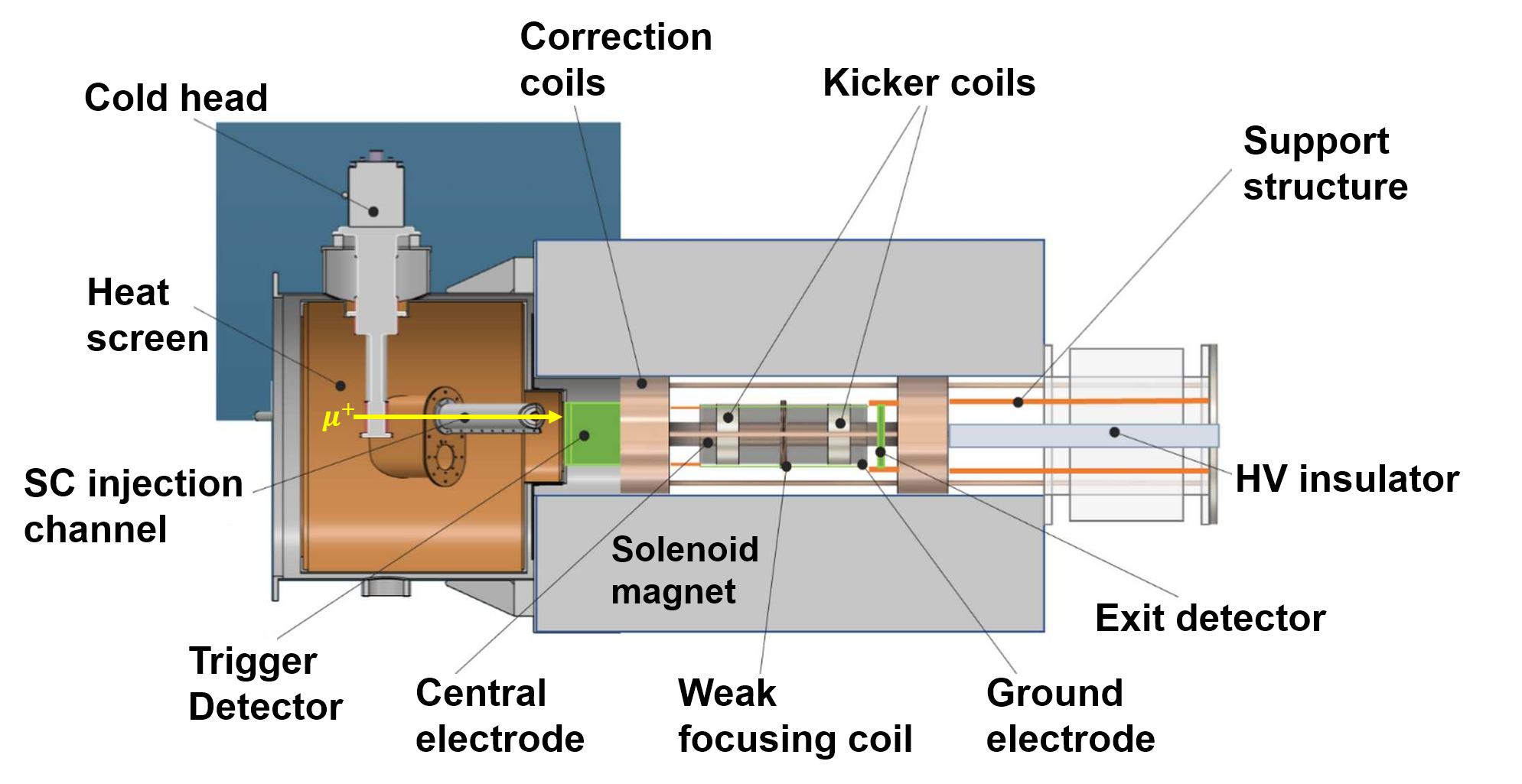}
\caption{Overview of the muEDM experiment at PSI. Displayed is the compact superconducting solenoid along with the phase-I experimental setup for the search for the muon EDM. The warm bore of the solenoid has an inner diameter of 200\,mm and an outer diameter of approximately 1000\,mm. Muons are injected into the solenoid through the superconducting (SC) injection channel before being detected by the trigger detector.}
\label{fig:Phase1Schematic}
\end{figure}

The PSI muEDM experiment utilizes the innovative frozen-spin technique~\cite{Farley:2003wt}, where muons are stored in a compact solenoid with a radial electric field, canceling the muon $g-2$ anomalous spin precession. This setup minimizes disturbance from the much stronger magnetic moment and allows precise measurement of spin rotation due to a potential EDM. In the experiment, a surface muon beam is injected into a compact PSC solenoid through a narrow superconducting (SC) channel~\cite{muEDM:2023mtc}, as depicted in Fig.~\ref{fig:Phase1Schematic}. Upon exiting the SC channel, the muon will be detected by a trigger detector~\cite{Wong:2024vmo}. For muon trajectories that match the storage phase space of the solenoid, a trigger signal will be generated by the detector. This trigger signal is used to activate a pulsed magnetic coil in the central region of the solenoid to deflect the trajectory of the muon into a stable orbit. Muons in the stable orbit will experience a radial E-field that cancels out the anomalous spin precession thus fulfilling the ``frozen-spin" condition~\cite{muEDM:2024bri} of the experiment. The muon EDM measurement can then proceed by measuring the upstream-downstream asymmetry of the decay positron count versus time, caused by the muon spin precession out of the orbital plane due to the muon EDM.

In this work, we report on the performance of a prototype muon trigger detector, developed to meet the stringent requirements of the muEDM experiment. This detector features plastic scintillators read out by silicon photomultipliers (SiPMs) and is designed to balance high detection efficiency with minimal beam perturbation. A detailed beam test was conducted using the $\pi$E1 beamline at PSI with 27.5~\text{MeV/c} muons to evaluate the detector’s performance under various configurations.

This paper is structured as follows: Section~\ref{sec:triggerdetector} describes the design and operational principles of the muon trigger detector. Section~\ref{sec:beamtestconfig} outlines the experimental setup, including the detector system, the data acquisition system, and the beamline configuration. Section~\ref{sec:measurement} presents the results of beam profile measurements, event characterization, and decay positron studies. Section~\ref{sec:sim_verification} validates these findings through Geant4-based simulations, and Section~\ref{sec:summary} concludes with a summary of the detector's performance and its implications for the PSI muEDM experiment.

\section{Muon trigger system}\label{sec:triggerdetector}

The muon trigger system~\cite{Wong:2024vmo} works in tandem with the magnetic pulse coil to store the muons for the muon EDM measurement. To maximize the sensitivity of the experiment, it must fulfill the following stringent requirements:
\begin{itemize}
    \item Detection efficiency of the incoming muon must be as high as possible;
    \item Perturbations to the muon beam trajectories must remain minimal in order to maintain a high storage efficiency;
    \item Rejection rate of non-storable muons must be as high as possible so that the DAQ rate is reduced to a manageable level;
    \item The time delay between the incoming muon and the trigger pulse generation must be as minimal as possible.
\end{itemize}

After investigating various detection options, we have decided on detectors based on plastic scintillators read out by SiPMs as the design is more flexible, and fast readouts have been demonstrated in various experiments. To understand how such an upstream detection system would affect the muon EDM measurement, we constructed a prototype muon trigger detector to address the top three requirements. The fourth requirement is addressed in a separate article using another version of the detector. The primary goal of the beam test was to measure the detector's response to different muon trajectories in the simplest scenario (without magnetic field) and to obtain light yield information from the plastic scintillators. 

\begin{figure}[htbp]
\centering 
\includegraphics[width=.99\textwidth]{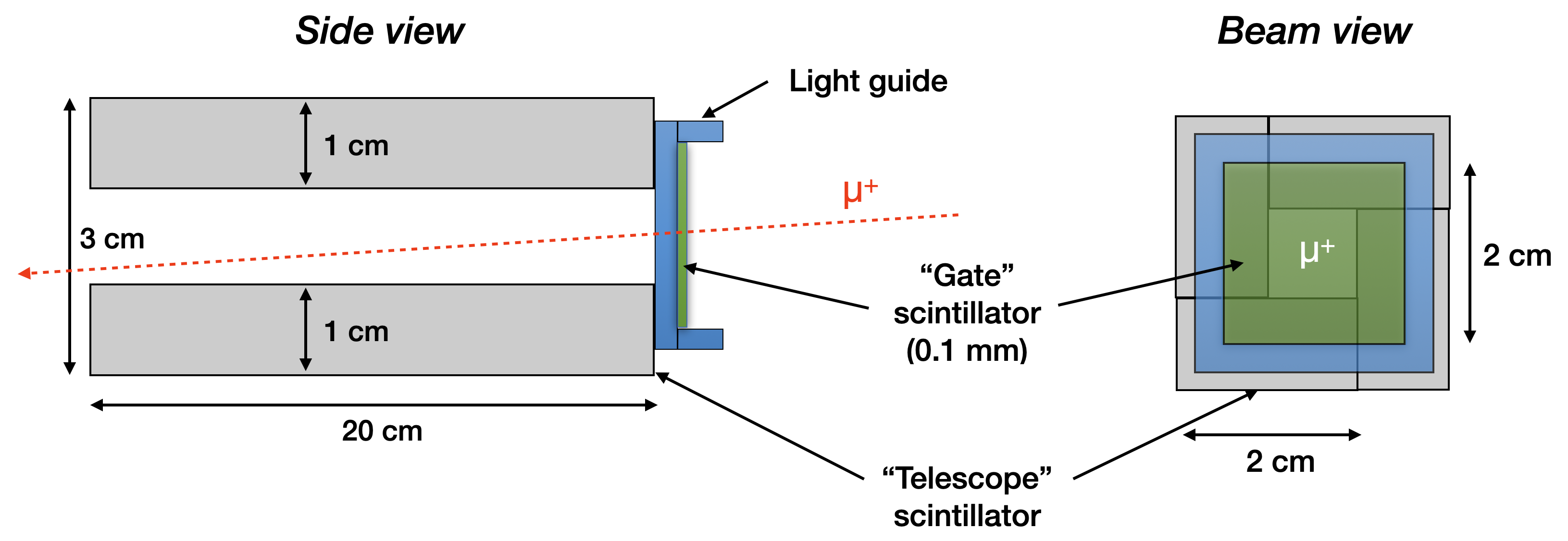}
\caption{Illustration of the prototype muon trigger detector. Muons that pass solely through the thin gate scintillator without triggering any telescope scintillator (anticoincidence condition) produce valid trigger events, indicating trajectories optimal for muon storage.}
\label{fig:triggerdetector}
\end{figure}

The muon trigger detector system, as shown in Fig.~\ref{fig:triggerdetector}, is composed of a \textit{gate} detector and a \textit{telescope} detector. If a muon passes through the gate detector without hitting the telescope sidewalls, it is potentially within the acceptance phase space and may be stored in a closed orbit. This means that in the case that both detectors observe a signal within the coincidence window, no trigger will be produced.

\begin{figure}
    \centering
    \includegraphics[width=.95\textwidth]{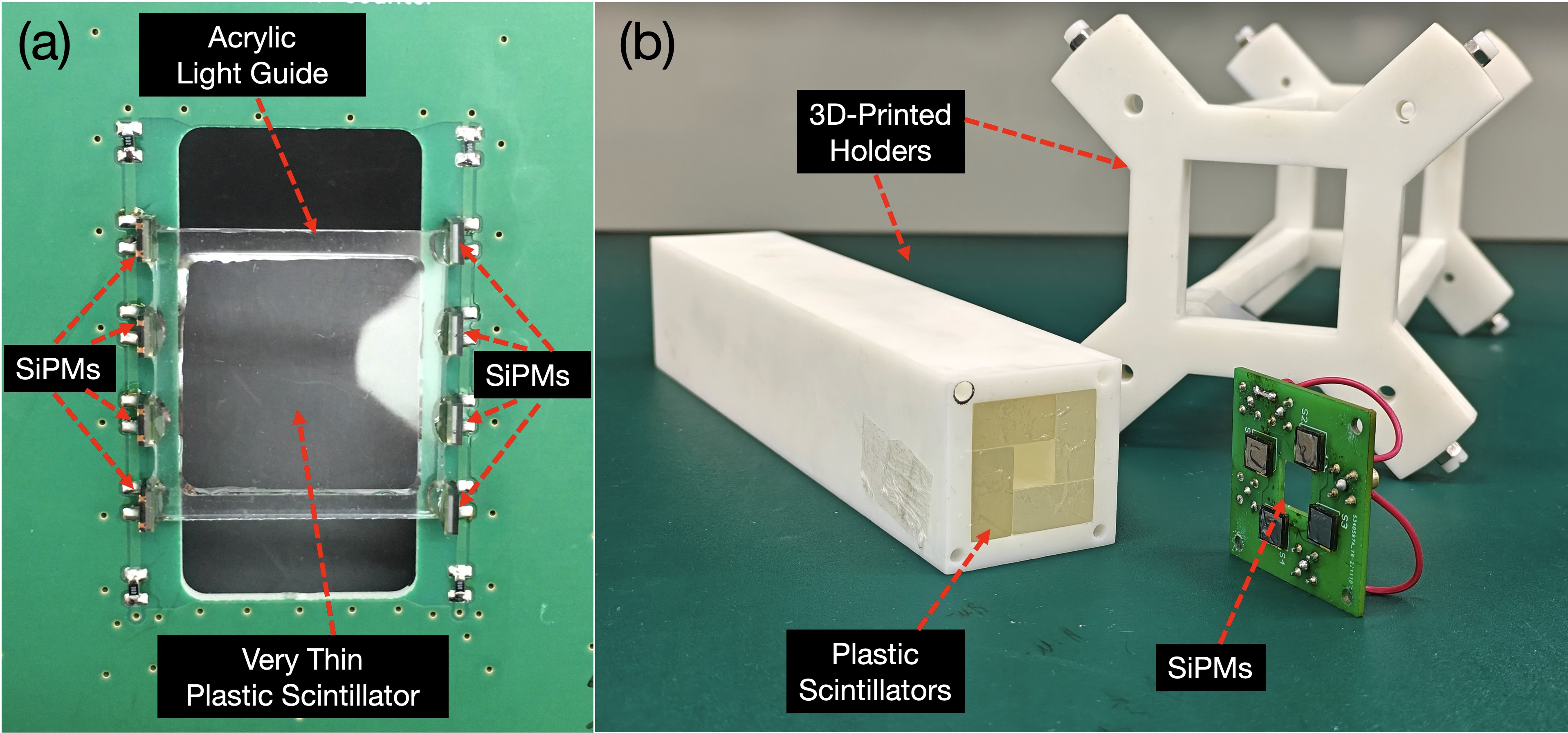}
    \caption{(a) Detailed view of the gate detector component. A thin BC-400 plastic scintillator (100\,$\mu$m thick, $20 \times 20$\,mm$^2$ area) is coupled to an acrylic light guide frame, read out by eight Hamamatsu S13360-1350 SiPMs positioned around its perimeter. (b) The four telescope scintillators are symmetrically and compactly arranged into a 3D-printed holder.}
    \label{fig:gatedetector}
\end{figure}

The gate detector~\cite{Stager:2023lgy} is a very thin \qtyproduct{20 x 20 x 0.1}{mm} BC400\footnote{LUXIUM Solutions, BC-400, BC-404, BC-408, BC-412, BC-416 Premium Plastic Scintillators: \href{https://luxiumsolutions.com/radiation-detection-scintillators/plastic-scintillators/bc400-bc404-bc408-bc412-bc416}{https://luxiumsolutions.com/radiation-detection-scintillators/plastic-scintillators/bc400-bc404-bc408-bc412-bc416}} plastic scintillator tile attached to a light guide frame of \qtyproduct{25 x 25}{mm} cross section. Eight Hamamatsu S13360-1350 SiPMs with an effective photosensitive area of $1.3 \times 1.3$\,mm$^2$ are attached to the sides of the light guide as shown in Fig.~\ref{fig:gatedetector} (a). The output signals of the eight SiPMs are combined into a single readout channel by an electronic board. A thickness of 100\,$\mu$m was selected based on a simulation study to minimize multiple Coulomb scattering (with the scattering angle restricted to around 5 degrees) while ensuring a sufficient number of detected photoelectrons in the SiPMs (averaging around 10 photoelectrons generated per SiPM). 

The telescope detector is composed of four GNKD HND-S2 scintillator bars, with the size of \qtyproduct{20 x 10 x 200}{mm}, as shown in Fig.~\ref{fig:gatedetector} (b). These scintillators are placed in a rectangular, compact, and symmetrical 3D-printed holder. Photons produced in one scintillator bar are also transmitted to other scintillator bars through optical cross-talk. The NDL EQR15-6060 SiPM\footnote{Novel Device Laboratory, Devices, SiPM: \href{http://www.ndl-sipm.net/products.html\#devices}{http://www.ndl-sipm.net/products.html\#devices}} with an active area \qtyproduct{6 x 6}{mm} was optically coupled to the downstream end of each plastic scintillator using BC-630 optical grease. 

\section{Beam test configuration}\label{sec:beamtestconfig}

\subsection{Beam configuration}

The beam test measurements and analysis described herein are based on a two-week run using the $\pi$E1 beamline at PSI in Nov-Dec 2022. The beamline was tuned to transport muons at a momentum of \qty{27.5}{MeV/c}. A scintillator-based beam counter on a 3-axis movable platform was setup in the downstream. The downstream beamline optics were tuned to maximize the count rate, with the counter positioned at $z=0$\,mm (roughly 300\,mm from the last quadrupole of the $\pi$E1 beamline) and $z=246$\,mm respectively. All tests described in the following were performed using the \qty{27.5}{MeV/c} muon beam at the $\pi$E1 beamline of PSI. There are two different beam settings tuned by quadrupole focusing of the $\pi E1$ beamline: Beam Tune A, which focuses the muon beam close to the gate detector ($z=0$\,mm), and Beam Tune B, which focuses the muon beam close to the exit detector ($z=246$\,mm). 

\subsection{Experimental setup}

Prior to the performance measurement of the prototype muon trigger detector, several detectors were installed and used to measure the muon beam profile:
\begin{itemize}
    \item \textbf{Veto detector}: A scintillator tile, size \qtyproduct{80 x 80 x 5}{mm}, with a hole of diameter \qty{18}{mm} at the center. This detector serves as a selection aperture matching approximately the geometrical acceptance of the telescope. Only muons going through the hole are taken into account.
    \item \textbf{Exit detector}: Behind the telescope detector, a thick scintillating tile size \qtyproduct{80 x 80 x 0.2}{mm} is used as the exit trigger of the setup.
    \item \textbf{Beam profile scanner}: Downstream of the entire setup, a 2D SiPM-based beam profile monitor (SiMon) was installed to measure the exit muon beam profile and quantify muons outside the geometrical acceptance of the prototype muon trigger detector.
\end{itemize}

\begin{figure}[htbp]
\centering 
\includegraphics[width=.9\hsize]{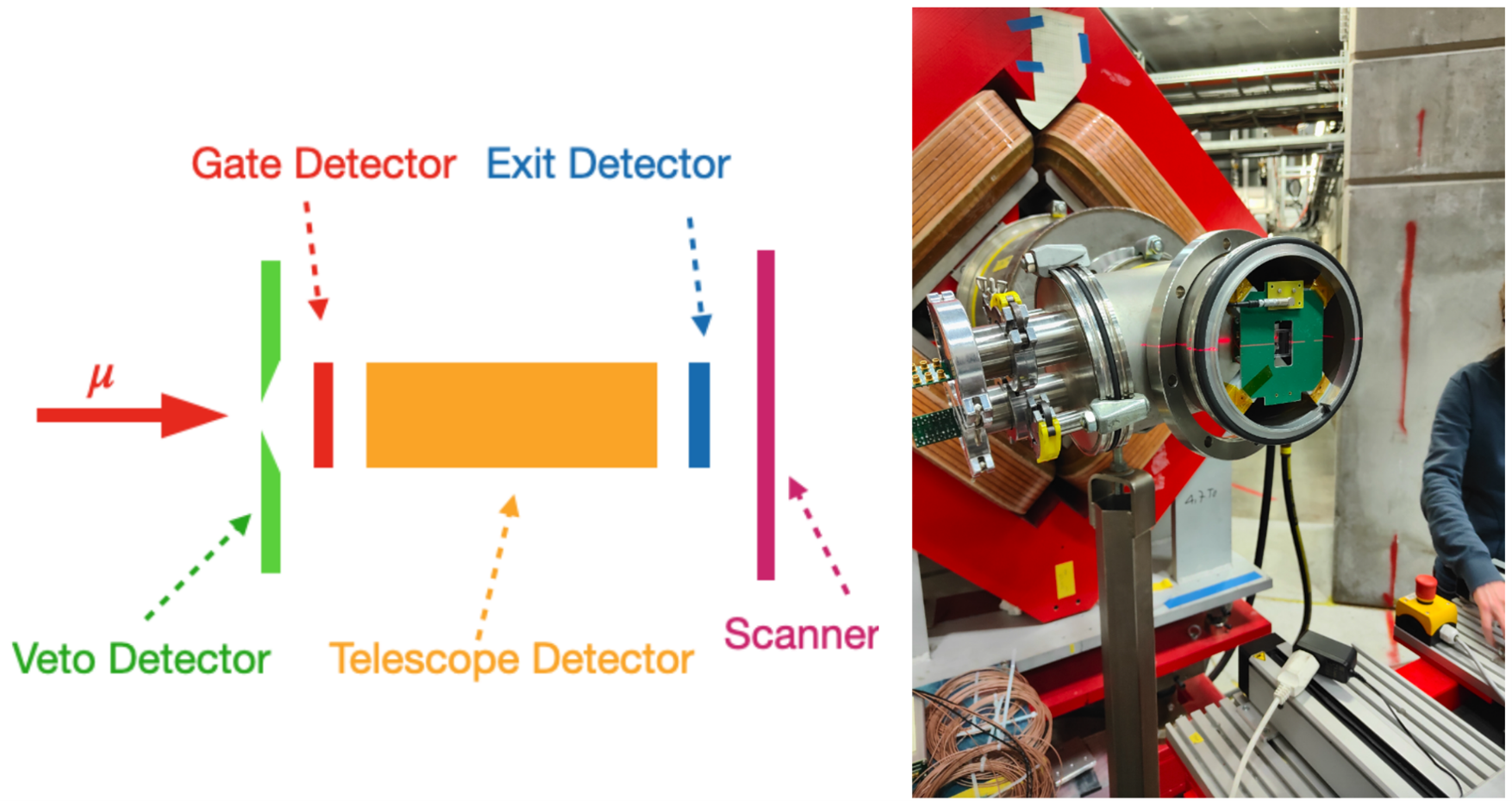}
\caption{Left: Schematic diagram showing the arrangement of auxiliary detectors used during beam testing, including the veto and exit detectors for event selection. Right: Photo showing the prototype muon trigger detector installed and aligned inside a vacuum tube (using a laser system) for testing at the PSI $\pi$E1 beamline.}
\label{fig:fullsetup_sketch}
\end{figure}

The experimental setup during the test beam is shown in Fig.~\ref{fig:fullsetup_sketch}. These auxiliary detectors ensured precise alignment and provided detailed measurements of the beam profile, helping to quantify the muon trajectories and evaluate the system’s performance.

\begin{figure}[htbp]
\centering 
\subfigure[Top view]{\includegraphics[width=.49\hsize]{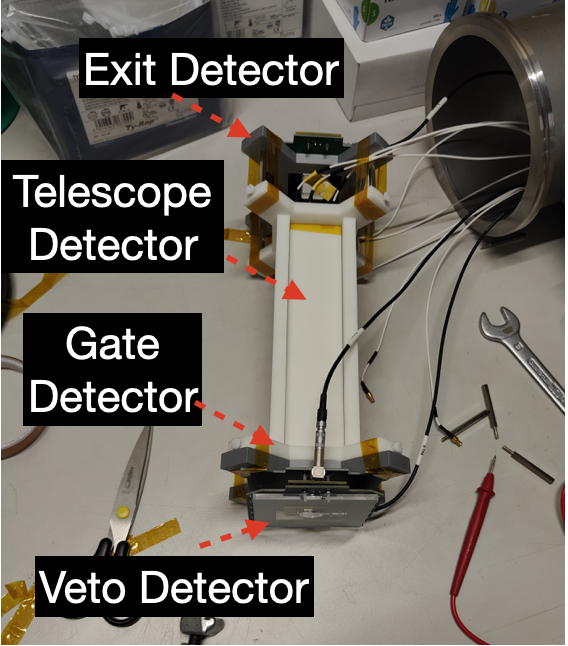}}
\subfigure[Front view]{\includegraphics[width=.49\hsize]{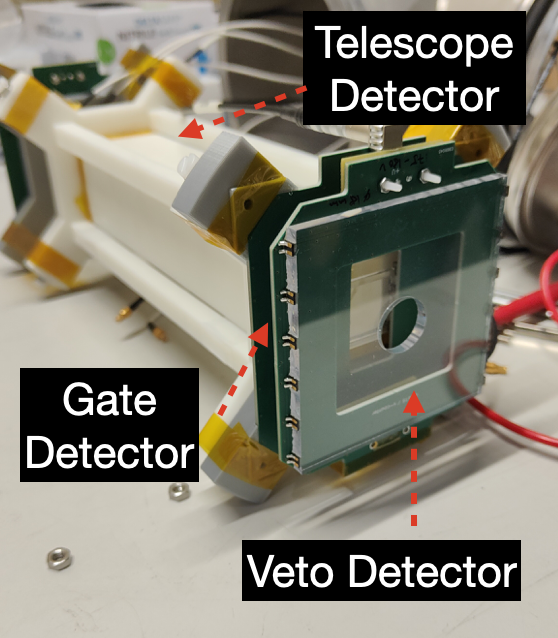}}
\caption{(a) Top view and (b) front view photographs of the fully assembled muon trigger detector setup, illustrating the arrangement of the gate and telescope detectors during beam tests.}
\label{fig:fullsetup_picture}
\end{figure}

A fully constructed muon trigger detector is shown in Fig.~\ref{fig:fullsetup_picture}. The setup was then installed in a vacuum flange to minimize the muon scattering with the air during the beam test. 

\subsection{Data Acquisition}

The Data Acquisition System (DAQ) employed in this experiment was the WaveDream Board (WDB)~\cite{Galli:2019nmv}, a 16-channel system developed at PSI for the MEG II experiment. The WDB integrates an advanced power supply and amplifier, allowing it to handle high-speed, high-precision signals essential for particle physics experiments. Its compact design and multi-channel capability make it well-suited for managing the complex data requirements of the prototype muon trigger detector.

\begin{figure}[htbp]
\centering
\includegraphics[width=.99\hsize]{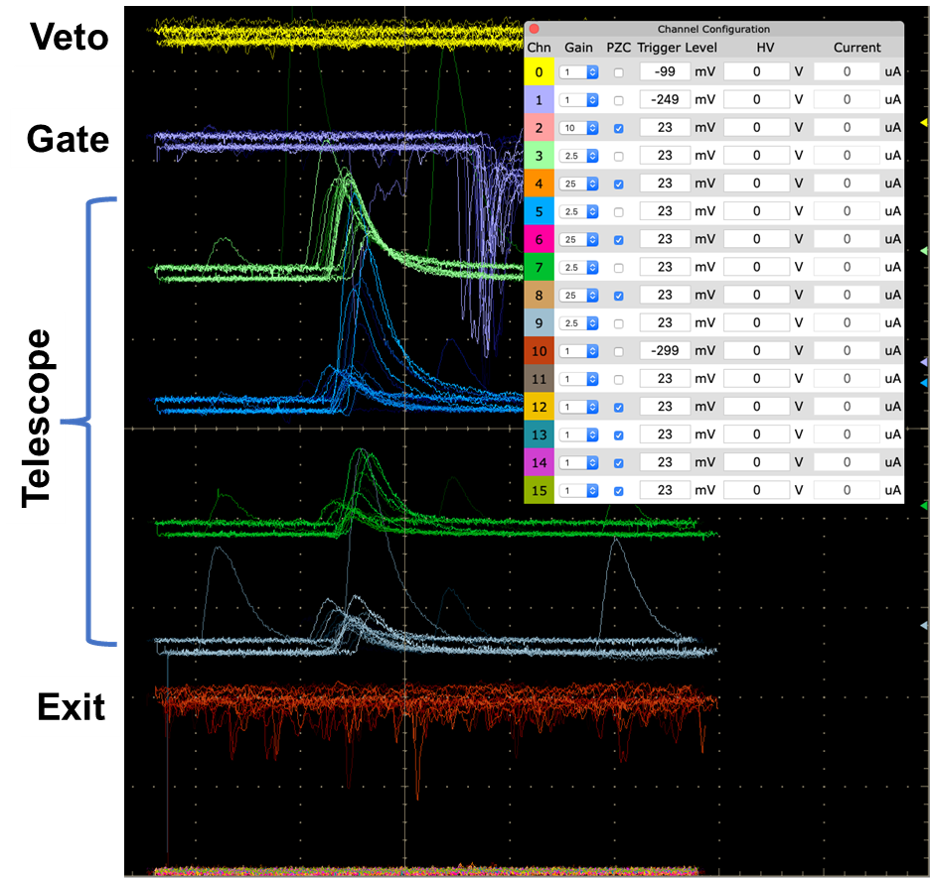}
\caption{A screenshot of the Graphical User Interface (GUI) for the WDB data acquisition system during the beam test. From top to bottom, the visible signals include SiPM signals (in persistence mode) from the veto, gate, telescope, and exit detectors. The individual trigger level of these signals is displayed in the top right corner of the image.}
\label{fig:waveDAQ}
\end{figure}

Figure~\ref{fig:waveDAQ} illustrates a screenshot from the WDB software, highlighting the signals captured during the beam test. The positive signals correspond to the four NDL SiPMs attached to the telescope detector, while the negative signals represent the veto detector, gate detector, and exit detector channels. The top right corner of Fig.~\ref{fig:waveDAQ} displays detailed information about each channel's trigger levels and gain settings, enabling precise control and optimization of the DAQ system.

During the beam test, approximately 0.7 million events were recorded under various trigger configurations as summarized in Tab.~\ref{tab:beamtest_datasets}. These triggers were designed to capture a comprehensive range of muon trajectories and interactions within the detector system. The configurations included:
\begin{enumerate}
    \item \textbf{Gate Self-Trigger}: Events were recorded when a muon hit the gate detector, independent of signals from other detectors.
    \item \textbf{Gate and Exit Coincidence}: Events were captured only when signals were registered simultaneously in both the gate and exit detectors, isolating muons that traversed the detector system.
    \item \textbf{Gate and Telescope Coincidence}: This setting recorded events where signals were simultaneously detected in the gate and telescope detectors, focusing on interactions within the telescope.
\end{enumerate}

\begin{table}[htbp]
\centering
\caption{Collected datasets during the beam test.}
\label{tab:beamtest_datasets}
\smallskip
\def\arraystretch{1.2}
\begin{tabular}{|c|c|c|}
\hline
Trigger Model & \multicolumn{2}{|c|}{Statistics}\\
\hline
& Beam Tune A & Beam Tune B\\ \hline
Gate Self-Trigger  & 100,000 & 150,000\\ \hline
Gate and Exit & 100,000 & 150,000\\ \hline
Gate and Telescope & 100,000 & 100,000\\ \hline
\end{tabular}
\end{table}

The DAQ system’s high sampling rate and low noise level ensured accurate signal capture and timing resolution, critical for differentiating between primary muon events and decay positrons. Its robust design also allowed seamless integration with the prototype detector setup, enabling efficient data collection during the two-week beam test.

\section{Experimental results}\label{sec:measurement}

\subsection{Beam profile extraction}

The beam profile measurements were crucial for understanding the spatial distribution and focusing properties of the 27.5 MeV/c muon beam at the $\pi$E1 beamline. These measurements informed the alignment and optimization of the prototype detector and facilitated the characterization of muon trajectories within the experimental setup.

Beam profiles were acquired using a combination of auxiliary detectors, including a veto detector with a central aperture, an exit detector, and a 2D beam profile monitor (SiMon) positioned downstream. The veto detector ensured that only muons within the expected acceptance region were considered for further analysis. The SiMon monitor provided high-resolution measurements of the transverse beam distribution at multiple positions along the beamline.

\begin{figure}[htbp]
\includegraphics[width=.99\hsize]{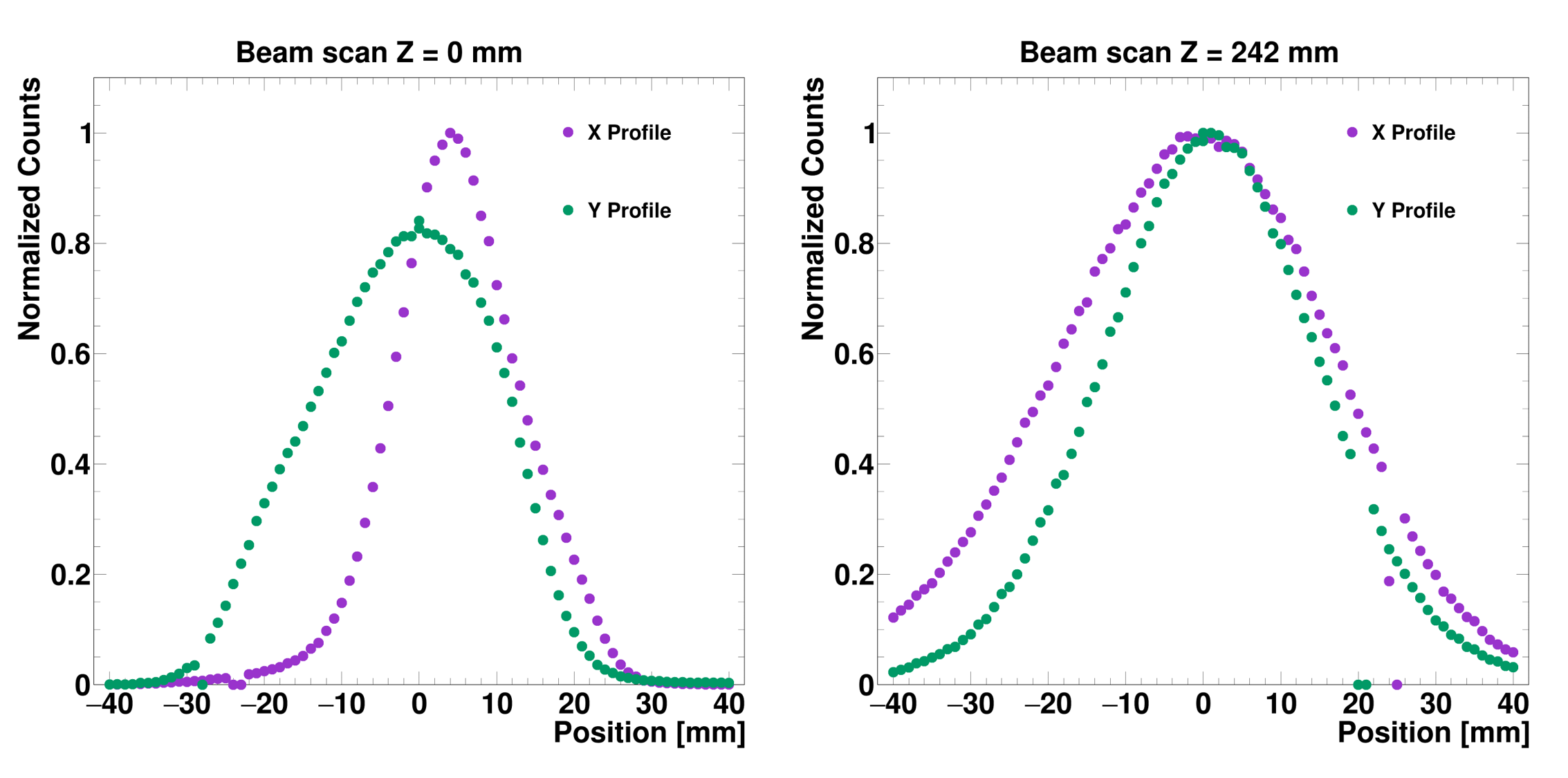}
\caption{Beam profile measurement for Beam Tune A where the beam is focused at $z=0$\,mm. Left: profile measured at $z = 0$\,mm; right: measured at $z = 246$\,mm.}
\label{fig:beamProfileZ0}
\end{figure}

\begin{figure}[htbp]
\includegraphics[width=.99\hsize]{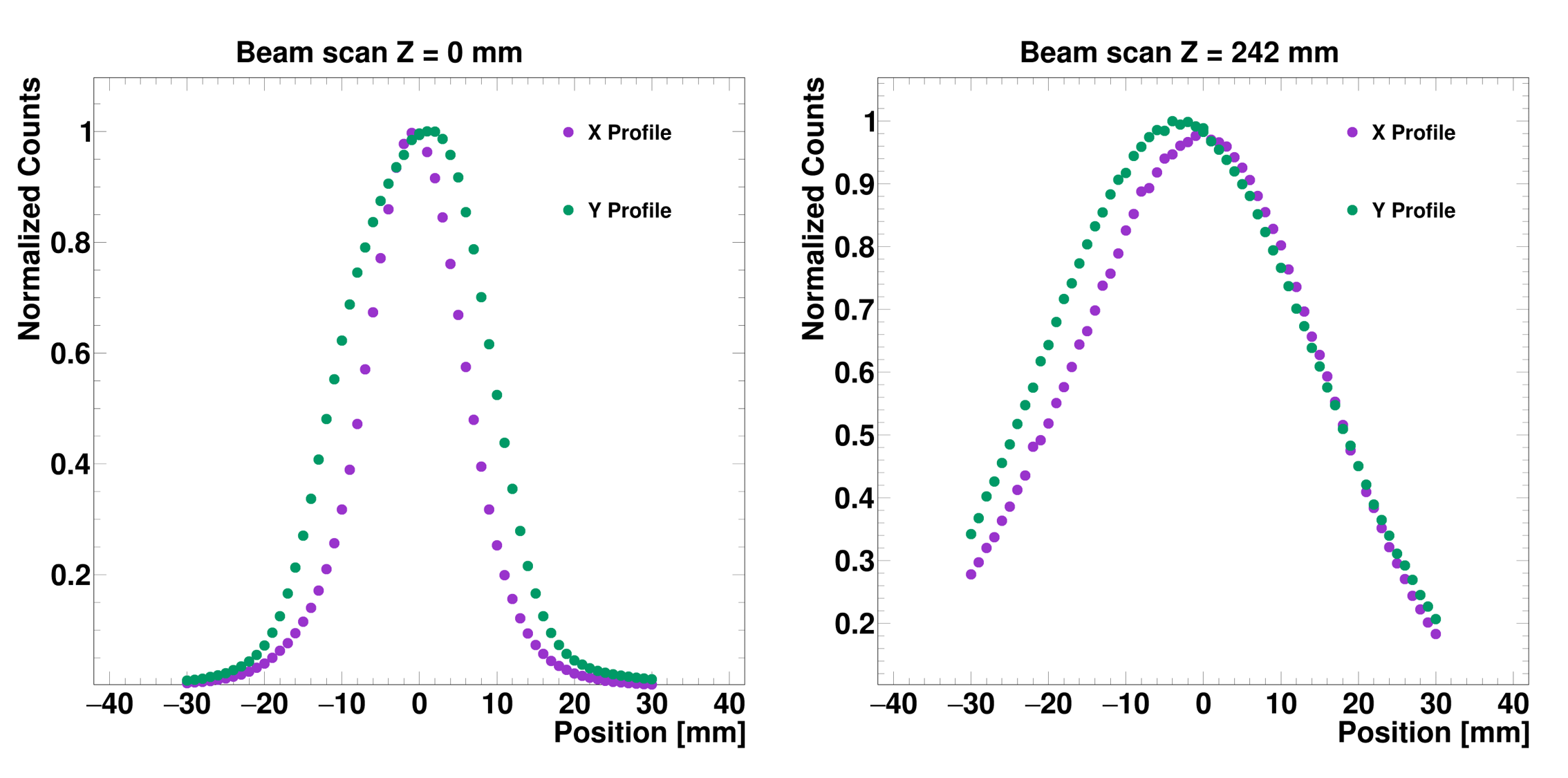}
\caption{Beam profile measurement for Beam Tune B where the beam is focused at $z=246$\,mm. Left: profile measured at $z = 0$\,mm; right: measured at $z = 246$\,mm.}
\label{fig:beamProfileZ246}
\end{figure}

Figures \ref{fig:beamProfileZ0} and \ref{fig:beamProfileZ246} illustrate the beam profiles for both tunes, measured at $z=0$~mm and $z=246$~mm, respectively. The profiles showed clear variations in beam size and intensity distribution, reflecting the effect of the quadrupole tuning parameters. For Beam Tune A, the beam was more collimated at $z=0$~mm, whereas Beam Tune B produced a broader distribution at the same position.

The beam sizes at different positions along the beamline were determined by fitting the measured profiles to Gaussian distributions. These sizes were corrected for Coulomb scattering effects in air to ensure accuracy. Using the corrected beam sizes, the beam optics formalism was applied to extract the Twiss parameters, including the horizontal and vertical beta functions and the emittance. These parameters provided a comprehensive description of the beam's phase space and were consistent with the expected beamline configuration.

Precise alignment of the detector components was achieved by cross-referencing the beam profiles with the known geometry of the setup. The alignment ensured that the beam trajectory intersected critical detector regions, such as the central aperture of the veto detector and the telescope scintillators. The measured profiles were also used as input for Geant4-based simulations, allowing direct comparison between measured and simulated beam distributions. As shown in Fig.~\ref{fig:beamModel}, the simulated phase space of the beam closely matched the experimental data, validating the accuracy of the beam model and the alignment procedure.

\subsection{Event characteristics and distributions}

To characterize the response of the prototype muon trigger detector, we analyzed the distribution of event topologies under different trigger models and beam tunes for the dataset collected during the beam test. The SiPM threshold was set to 4.5 photoelectrons (p.e.), effectively suppressing dark noise while reliably selecting muon events detected by the telescope. This threshold was applied uniformly during online data acquisition and offline analysis.

\begin{figure}[htbp]
\centering 
\subfigure[Gate detector self-trigger]{\includegraphics[width=.85\hsize]{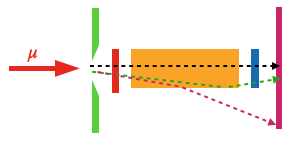}
\label{Gate Trigger}}
\qquad
\subfigure[Gate coincide with Exit]{\includegraphics[width=.85\hsize]{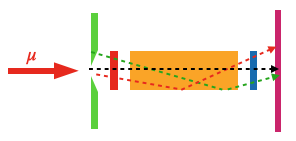}
\label{Gate coincide with Exit}}
\subfigure[Gate coincide with Telescope]{\includegraphics[width=.85\hsize]{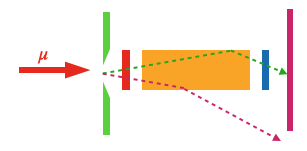}
\label{Gate coincide with Telescope}}
\caption{Visualization of different event topologies identified during beam testing under distinct trigger conditions: (a) Gate detector self-triggered events, (b) gate detector signals coincident with exit detector signals, and (c) gate detector signals coincident with telescope detector signals. Each scenario highlights characteristic muon interaction pathways through the experimental setup.}
\label{fig:event_topology}
\end{figure}

Event classification was performed based on the combination of signals from the veto, gate, telescope, and exit detectors. These configurations helped distinguish between muons that passed through the acceptance region, stopped in the telescope, or scattered outside the defined geometry. Table \ref{tab:event_topology_distribution} summarizes the measured event distributions for each trigger model, with representative topologies visualized in Fig.~\ref{fig:event_topology}.

For instance, the topology ``(!Veto)~\&~Telescope~\&~(!Exit)'' corresponds to events where the muon bypassed the veto detector (indicating passage through the central hole), produced a signal in the telescope detector, but did not reach the exit detector. Such events typically arise from muons stopping within the telescope or scattering outside the exit detector's acceptance.

We observed distinct event distributions for the three trigger models:
\begin{itemize}
    \item \textbf{Gate Self-Trigger}: Most of the muons (83.99\%–86.72\%) entering the detector did not trigger the veto (thus not coming from undesired directions or regions) and did not register a signal in the exit detector. This result suggests that a large portion of muons either stopped or scattered significantly within the telescope detector.
    \item \textbf{Gate and Exit Coincidence}: Most of the muons (76.00\%–76.80\%) entering the detector did not trigger the veto or register a signal in the telescope detector. This result suggests that a significant portion of muons passes through the detector as intended. This further confirms that the detector's veto system and geometry effectively reduce background and unintended muon trajectories.
    \item \textbf{Gate and Telescope Coincidence}: A small fraction of events exists for a triple coincidence among the gate, telescope, and exit detectors. This corresponds to muons scattering on the telescope detector and hitting the exit detector downstream. 
\end{itemize}

\begin{table*}[htbp]
\centering
\caption{Measured distribution for each trigger mode and event topology. In the table, ``!'' represents that no signal was recorded in the specified detectors, for example '!Veto' represents that no signal was registered in the veto detector, and ``\&'' represents a coincidence between two readouts.}
\label{tab:event_topology_distribution} 
\smallskip
\def\arraystretch{1.5}
\begin{tabular}{|c|c|c|c|}
\hline
Trigger Mode  & Event Topology & Beam Tune A & Beam Tune B \\ \hline
\multirow{3}{*}{Gate Self-Trigger}  & (!Veto)~\&~Telescope~\&~(!Exit) &  86.72\% &  83.99\%   \\ \cmidrule{2-4} 
                                    & (!Veto)~\&~Telescope~\&~Exit  &  0.75\%  &   1.17\%   \\ \cmidrule{2-4} 
                                    & (!Veto)~\&~(!Telescope)~\&~Exit              &  2.78\%  &   4.78\%   \\ \hline
\multirow{3}{*}{Gate and Exit}      & Veto            &   4.47\%    &   4.86\%        \\ \cmidrule{2-4} 
                                    & (!Veto)~\&~Telescope &  19.53\%    &   18.34\%      \\ \cmidrule{2-4} 
                                    & (!Veto)~\&~!Telescope             &  76.00\%    &  76.80\%        \\ \hline
\multirow{2}{*}{Gate and Telescope} & (!Veto)~\&~(!Exit)  &  87.71\%    &    84.65\%      \\ \cmidrule{2-4} 
                                    & (!Veto)~\&~Exit   &   3.59\%    &  5.77\%          \\ \hline
\end{tabular}
\end{table*}

The spatial distribution of events revealed beam trajectory variations and their impact on muon interactions. Events were more concentrated near the central region under Beam Tune A, while Beam Tune B produced a broader distribution near the exit detector. These observations were consistent with the measured beam profiles (Fig.~\ref{fig:beamProfileZ0} and~\ref{fig:beamProfileZ246}) and provided additional insight into beam alignment and focusing effects.

\subsection{Decay positron signals}

\begin{figure}[htbp]
\centering 
\subfigure[]
{\includegraphics[width=0.9\textwidth]{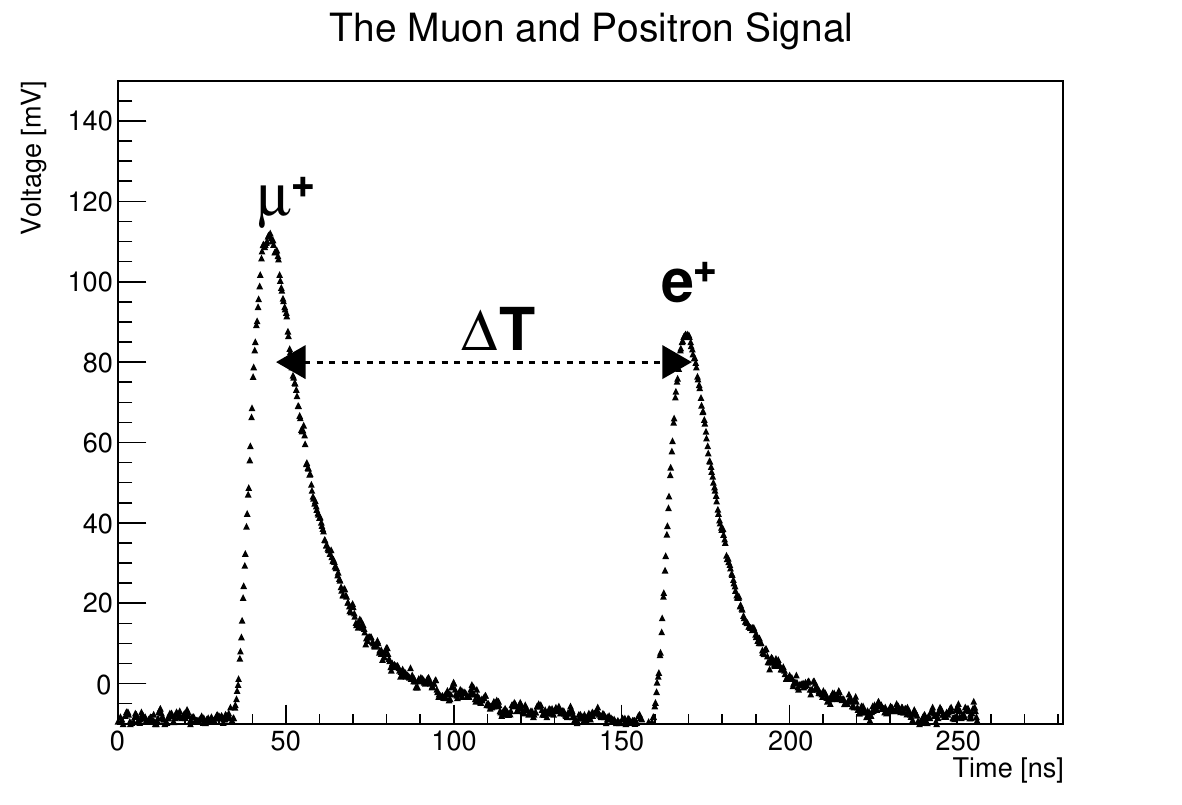}
\label{fig:Two peaks appeared in a single event.}}
\qquad
\subfigure[]
{\includegraphics[width=0.9\textwidth]{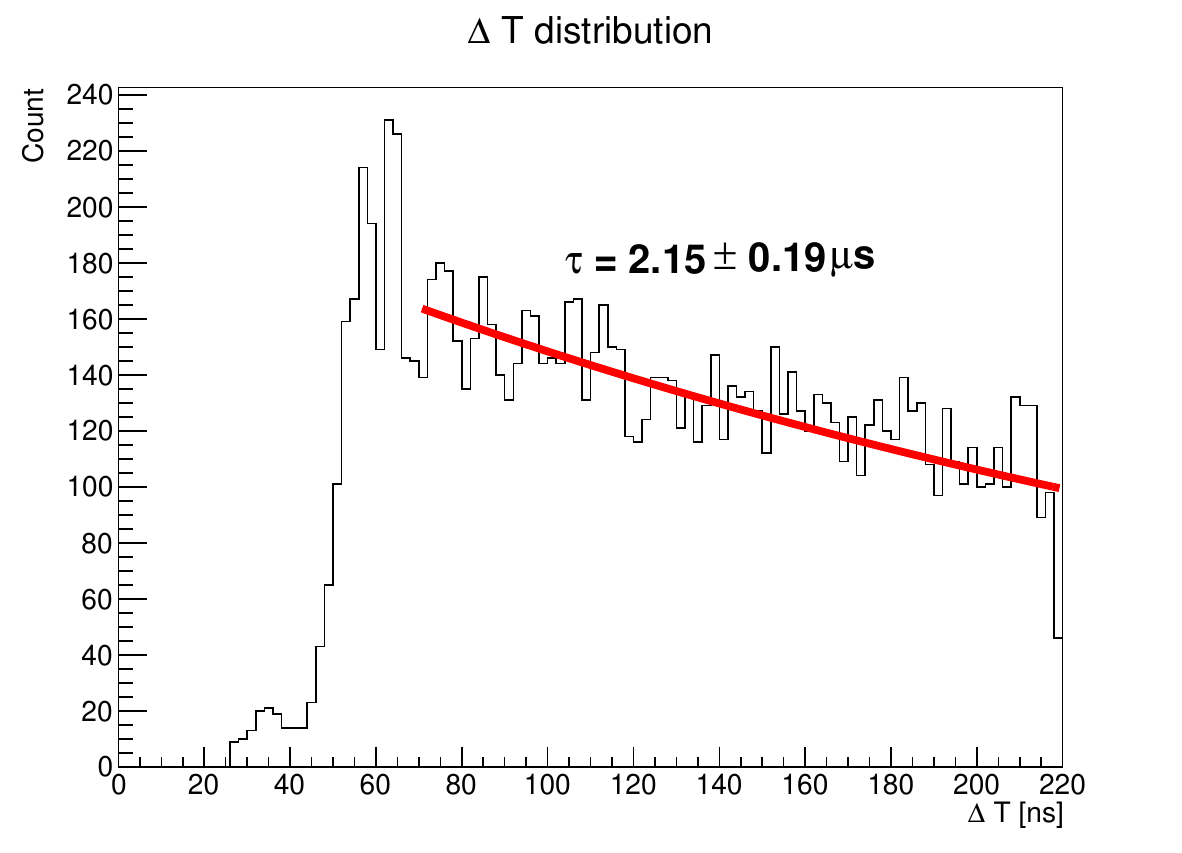}
\label{fig:Time interval distribution for Channel 5}}
\caption{(a) Typical waveform showing a double-peak signal corresponding to a muon and its decay positron within the telescope scintillator. (b) Histogram and exponential fit of the time intervals ($\Delta T$) between the two signals, confirming the muon decay signature with a fitted lifetime of $2.15 \pm 0.19\,\mu$s, consistent with the known muon lifetime.}
\label{fig:muon_decay_events} 
\end{figure}

During offline waveform reconstruction, two distinct peaks were identified in some events, as shown in Fig.~\ref{fig:muon_decay_events}. Given the muon momentum of \SI{27.5}{MeV/c}, it is expected that muons would stop within the telescope detector scintillator, allowing their decay positrons to be detected by the same scintillator. To confirm that the later peak arises from a decay positron, we analyzed the time intervals, $\Delta T$, between the two peaks.

The distribution of $\Delta T$ was fitted with an exponential decay function $N_{0} e^{-\Delta T/\tau}$, where $N_{0}$ is the normalization, and $\tau$ the muon lifetime. The results for one of the scintillators in the telescope detector are shown in Fig.~\ref{fig:muon_decay_events} (b). The fitted exponential decay constant yielded a lifetime of \(2.15(19)\,\mu s\), which is in good agreement with the measured muon lifetime of $2.1969803(22)\,\mu s$~\cite{MuLan:2012sih}. Despite the large uncertainty, the analysis validates the detector's capability to capture and resolve decay positron signals with high temporal accuracy.

\subsection{Detector response analysis}

The beam test provided critical insights into the optical response characteristics of the prototype detector, particularly regarding scintillator light yield and SiPM response. SiPMs were calibrated using single-photon spectrum analysis. During calibration, signals from each SiPM were recorded under low-light conditions, enabling the measurement of single-photon peaks. Each of these peaks corresponds precisely to integer numbers of detected photons. By fitting these photon-count peaks with a multi-Gaussian distribution, the spacing between consecutive peaks—corresponding to the charge produced by a single photo-electron was accurately determined. This calibration procedure allowed for converting raw detector signals into the number of photo-electrons, thus ensuring accurate and consistent photon-counting performance across measurements.

\begin{figure}[htbp]
\centering 
\includegraphics[width=0.99\textwidth]{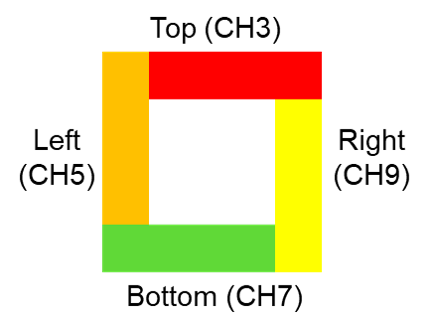}
\caption{Diagram illustrating the arrangement and labeling of the four scintillator bars within the telescope detector, each coupled individually to an SiPM for independent optical readout.}
\label{fig:telCartoon}
\end{figure}

The telescope detector consisted of four scintillator bars, each connected individually to an SiPM, as shown in Fig.~\ref{fig:telCartoon}. As muons passed through a scintillator bar, scintillation photons were produced and subsequently collected by the SiPM mounted on the same bar. Furthermore, optical cross-talk led to photon collection by SiPMs connected to neighboring scintillators.

Measurements indicated an average yield exceeding 300 detected photons per muon event by the directly impacted scintillator’s SiPM. Adjacent scintillators collected approximately 100 to 120 photons, while the opposite scintillator recorded around 50 photons per event. These photon yields align well with theoretical expectations derived from the optical properties of the scintillator materials and the geometrical configuration of the detector. The observed photon yield was more than sufficient for generating robust anti-coincidence signals, crucial for the requirement of the experiment.

\begin{figure*}[htbp]
\centering 
\subfigure[Expected optical photon number correlation of adjacent channels (CH9 vs CH3)]{\includegraphics[width=0.45\textwidth]{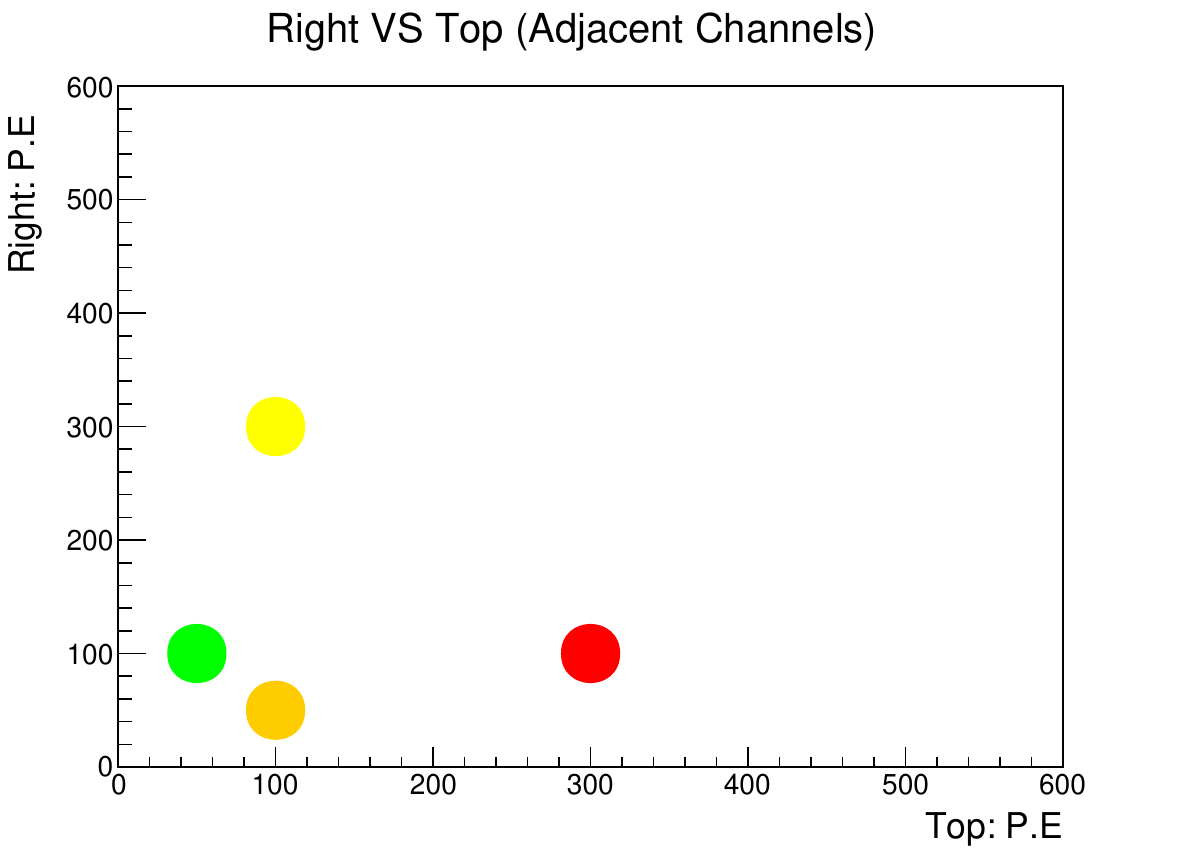}
\label{Expected distribution for CH3 versus CH9}}
\qquad
\subfigure[Expected correlation of opposite channels (CH7 vs CH3)]{\includegraphics[width=0.45\textwidth]{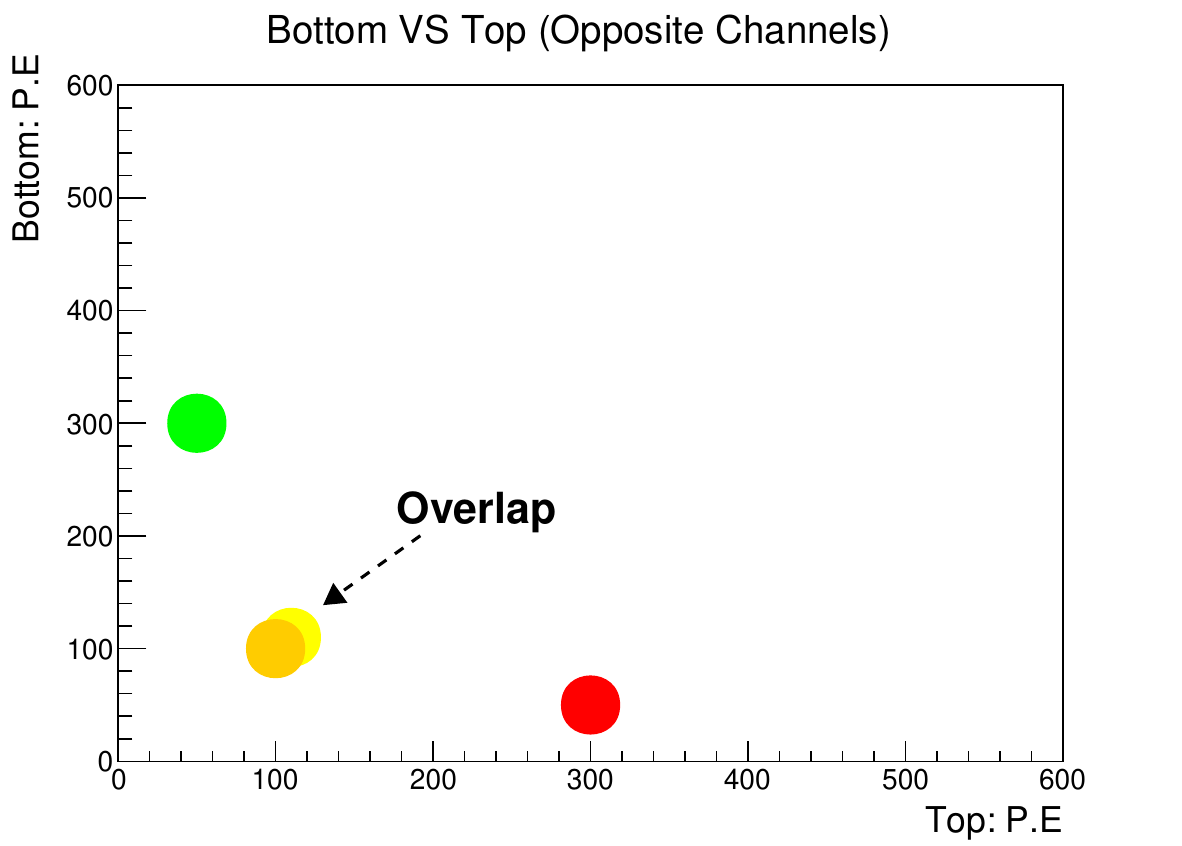}
\label{Expected distribution for CH3 versus CH7}}
\qquad
\subfigure[Measured optical photon number correlation of adjacent channels (CH9 vs CH3)]{\includegraphics[width=0.45\textwidth]{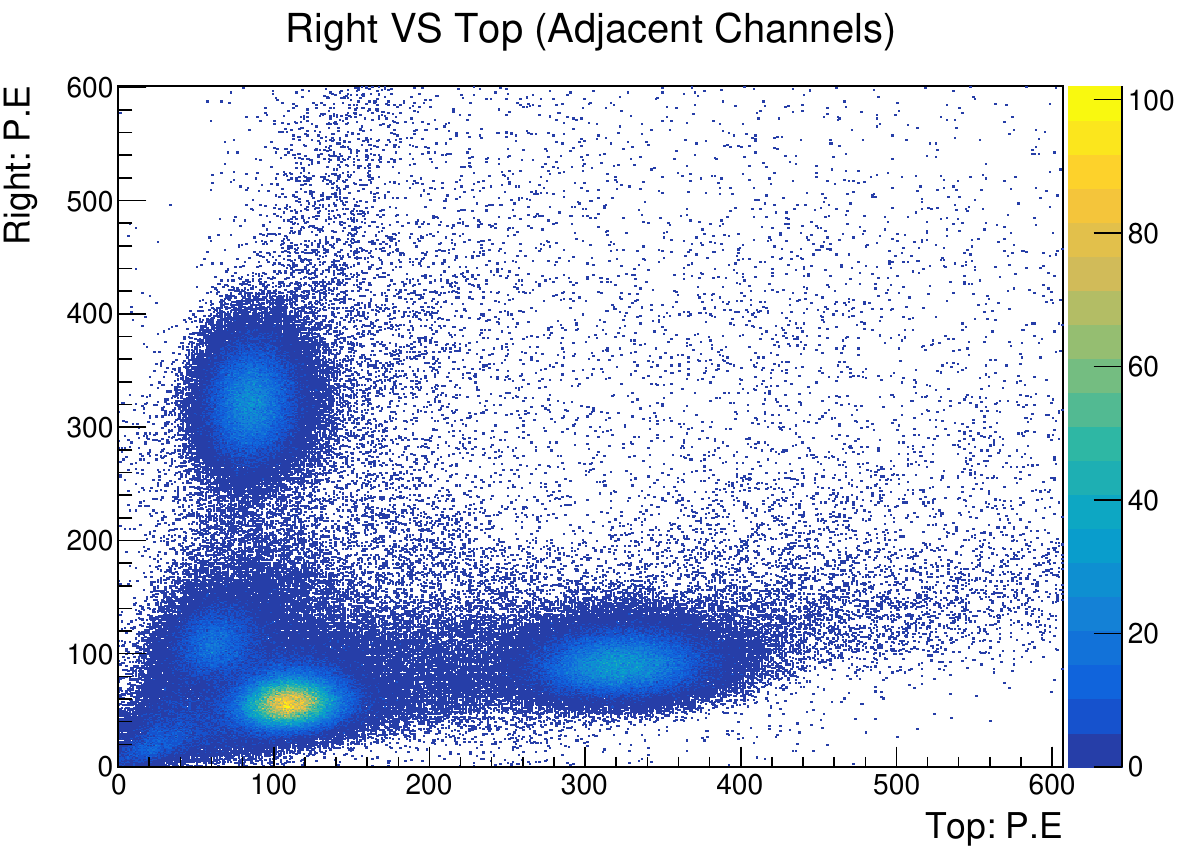}
\label{Experimental distribution for CH3 versus CH9}}
\qquad
\subfigure[Measured optical photon number correlation of opposite channels (CH7 vs CH3)]{\includegraphics[width=0.45\textwidth]{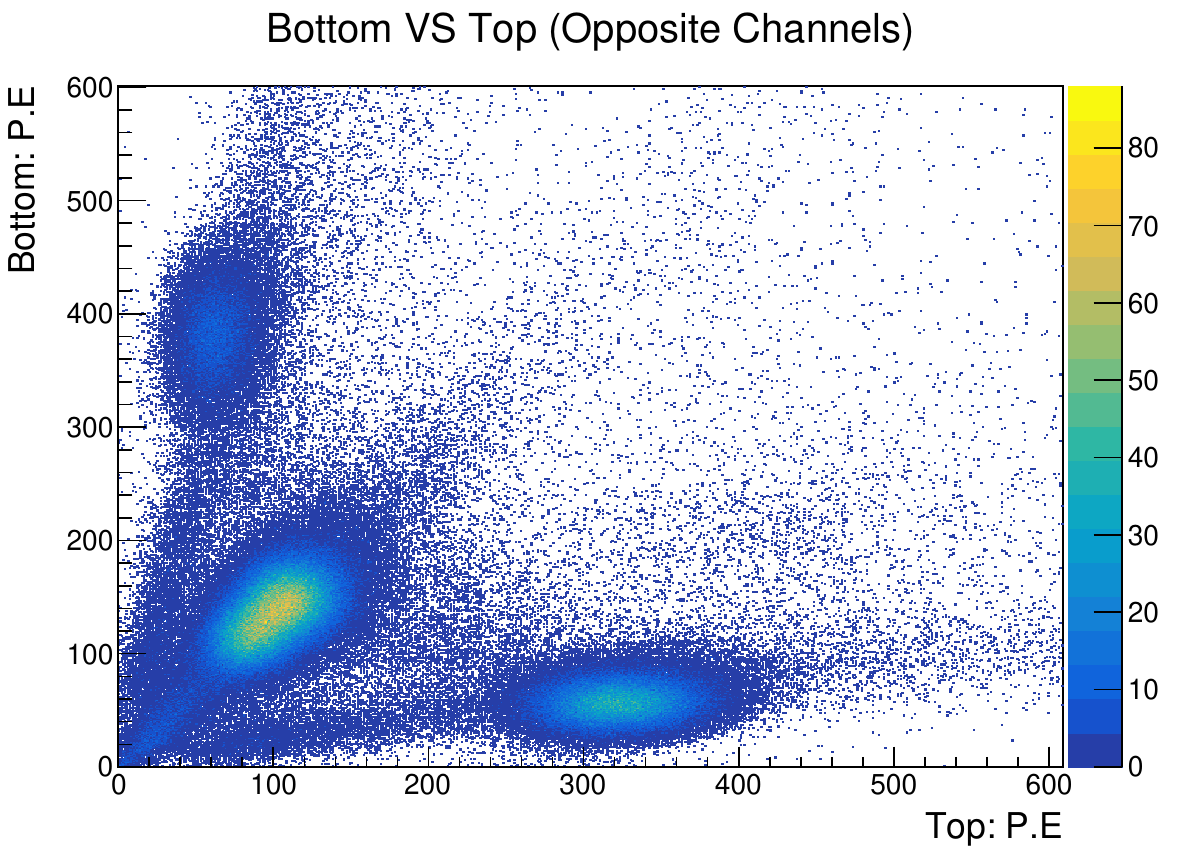}
\label{Experimental distribution for CH3 versus CH7}}
\qquad
\caption{Correlation plots of SiPM photo-electron counts collected by adjacent and opposite scintillator bars in the telescope detector. Panels (a) and (b) show expectations of photon distribution due to optical cross-talk for muon-only events. Panels (c) and (d) present experimentally measured photon distributions, validating the cross-talk behavior and optical modeling accuracy.}
\label{fig:photon_distribution}
\end{figure*}

Photon yield distributions were analyzed to assess symmetry and quantify cross-talk between scintillators. Correlation plots for adjacent and opposite scintillator pairs demonstrated distinct photon-sharing patterns. Specifically, correlations between the directly hit scintillator and adjacent scintillators revealed multiple peaks, highlighting expected photon distribution characteristics due to optical cross-talk. In contrast, correlations involving the opposite scintillator displayed a single, lower-intensity peak indicative of minimal cross-talk. These correlations, illustrated in Figs.~\ref{fig:photon_distribution}(c) and \ref{fig:photon_distribution}(d), closely matched predictions from optical models.

\section{Simulation Verification}\label{sec:sim_verification}

To validate the experimental results, Geant4-based simulations were conducted using the musrSim framework~\cite{Sedlak:2012}. This section details the methodology, setup, and comparative analysis between simulations and experimental data, focusing on two critical aspects: muon beam phase space distribution and event topology. The simulations aimed to reproduce the experimental conditions, including beam dynamics, detector geometry, and optical photon transport, to confirm the detector’s design accuracy.

\subsection{Muon Beam Phase Space Distribution}

\begin{figure*}[htbp]
\centering 
\subfigure[]{\includegraphics[width=0.45\textwidth]{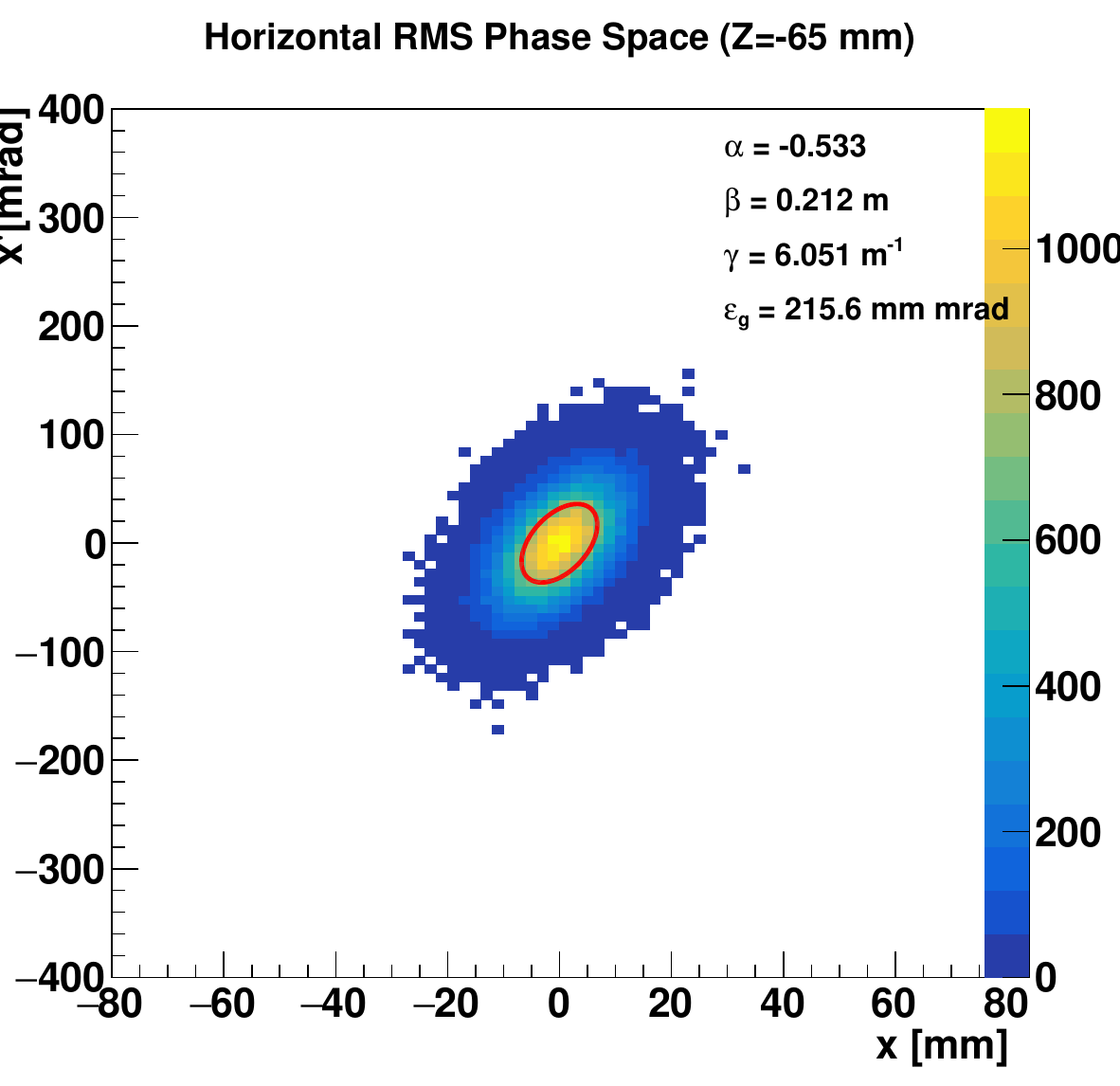}}
\qquad
\subfigure[]{\includegraphics[width=0.45\textwidth]{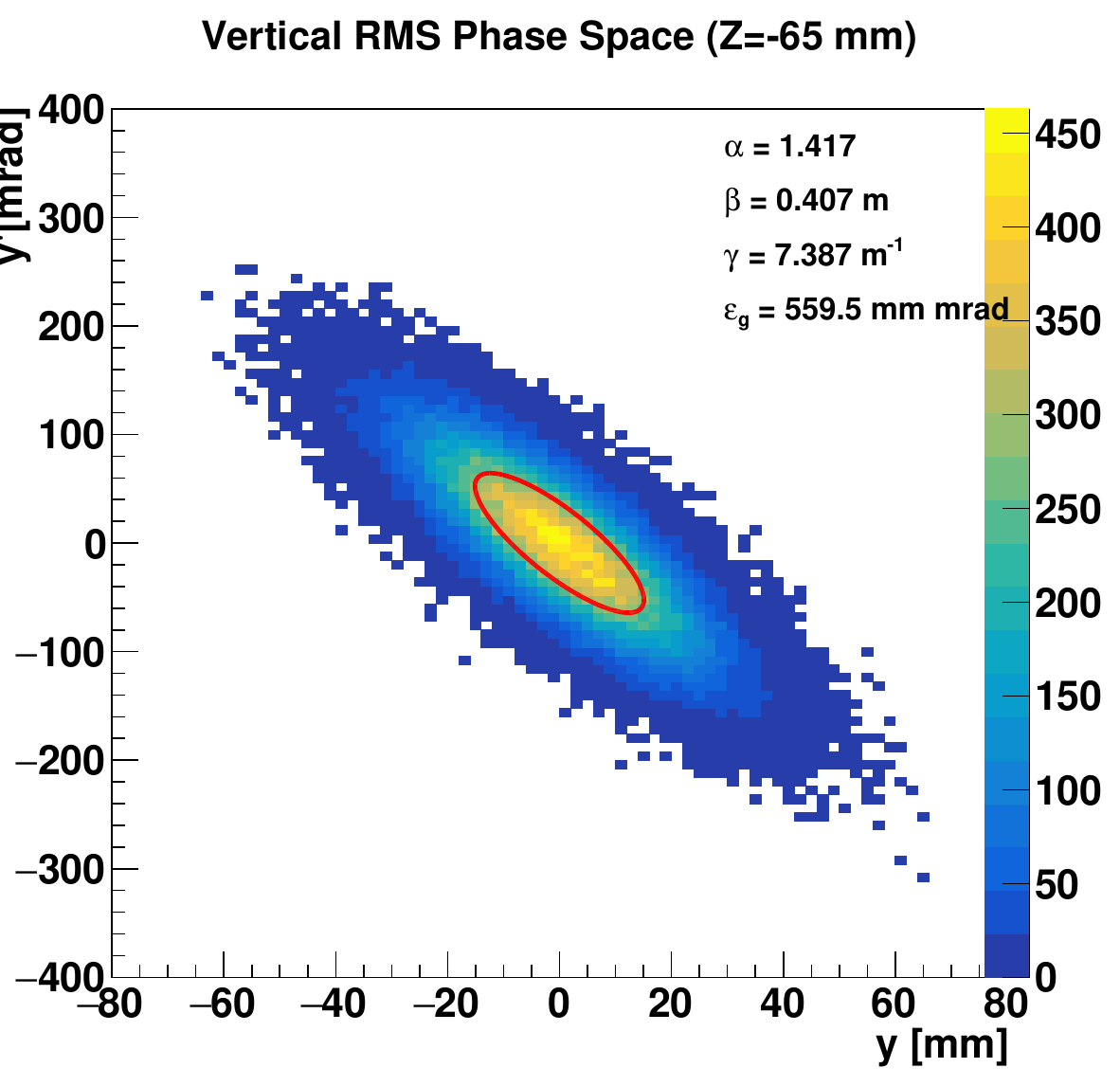}}
\qquad
\subfigure[]{\includegraphics[width=0.45\textwidth]{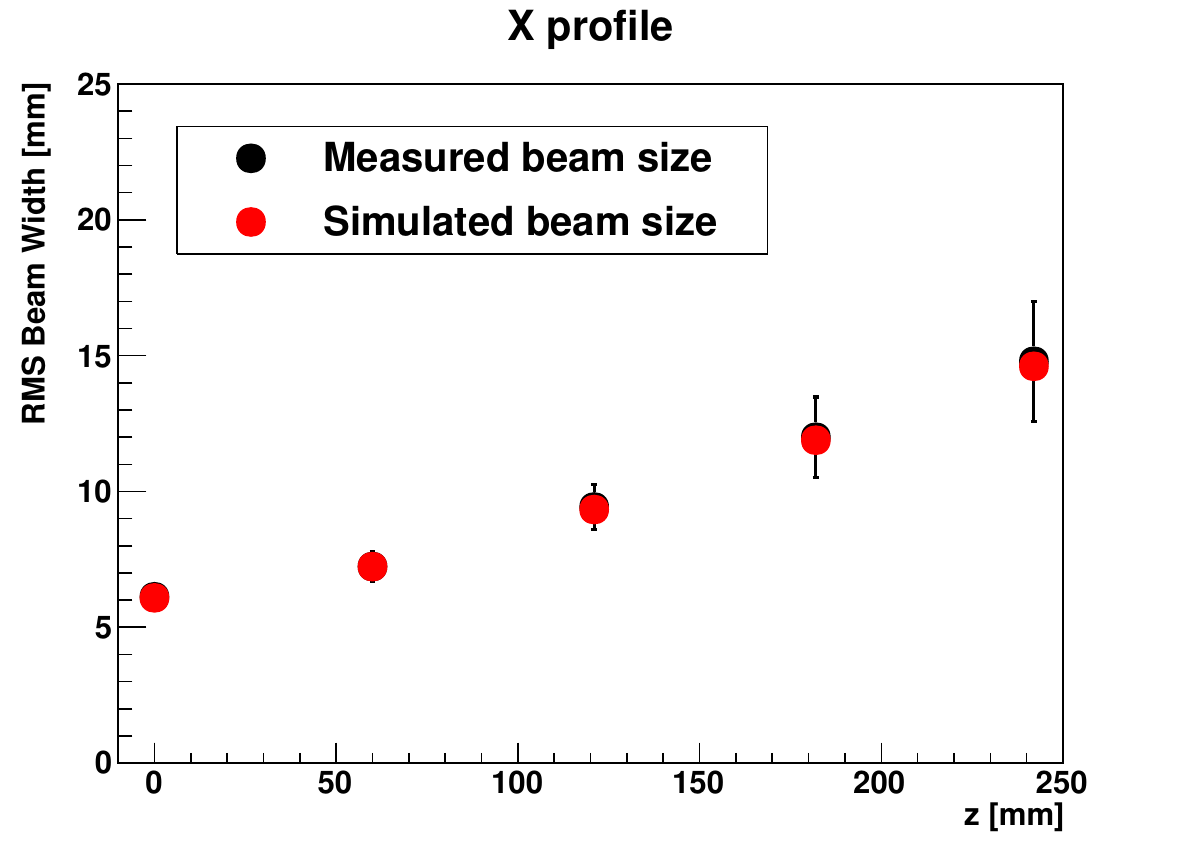}}
\qquad
\subfigure[]{\includegraphics[width=0.45\textwidth]{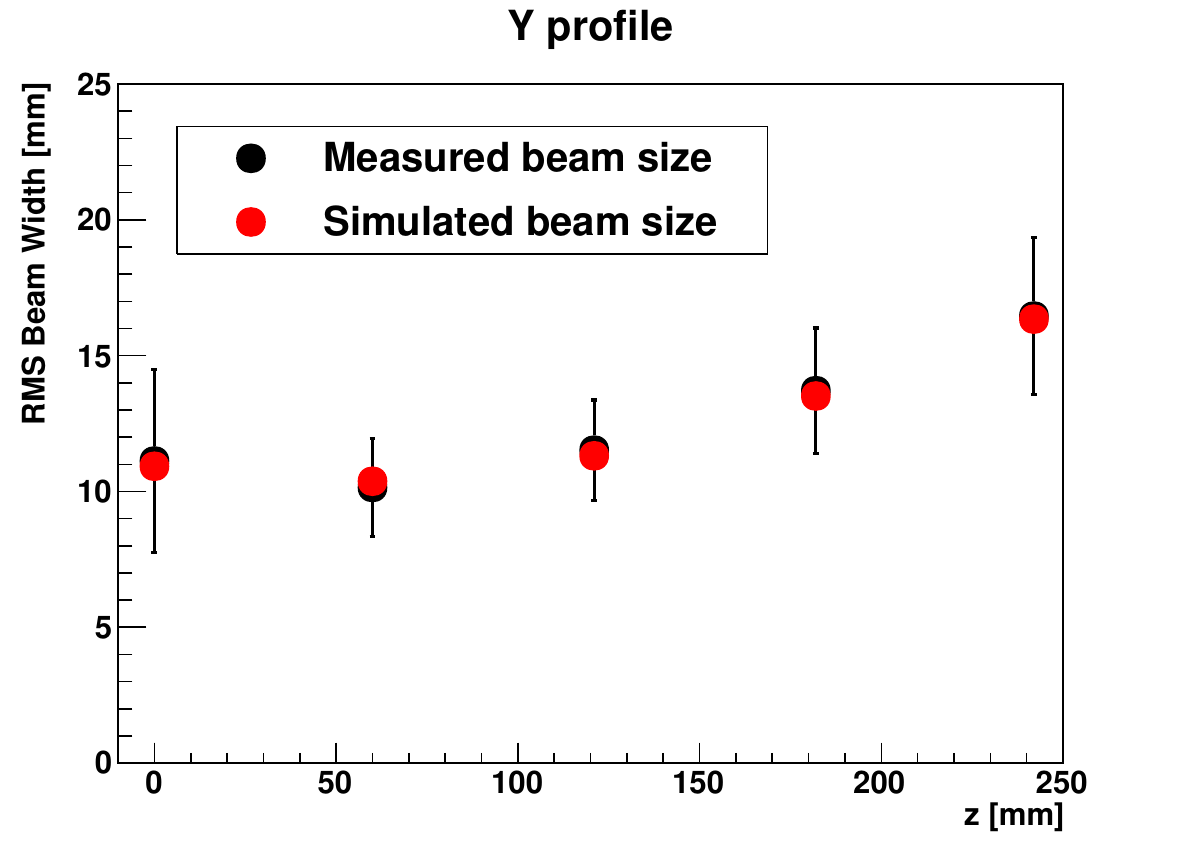}}
\caption{Comparison of the measured and simulated muon beam parameters. The top panels display the simulated muon beam phase space distributions (horizontal and vertical) at $z = –65$\,mm. The red ellipse represents the phase space ellipse derived from the measured Twiss parameters. The bottom panels compare the measured and simulated beam sizes (RMS widths) along the beamline, confirming that the simulations accurately reproduce the experimental beam conditions.} \label{fig:beamModel}
\end{figure*}

The muon beam’s phase space and trajectory were simulated in Geant4 using the miniScatter package~\cite{Sjobak:2019bmq}, a Geant4 wrapper optimized for users at particle accelerator facilities. Twiss parameters ($\alpha$, $\beta$, and $\gamma$), which describe the beam’s transverse emittance (a measure of beam spread) and focusing properties, were derived from Gaussian fits to the measured beam profiles at $z=0$ and $z=246$\,mm (Figs.~\ref{fig:beamProfileZ0} and \ref{fig:beamProfileZ246}), and implemented in the simulation. Figure~\ref{fig:beamModel} illustrates the comparison between the simulation and measurement of the muon beam phase space at the end of the $\pi$E1 beamline ($z=-65$\,mm) and the beam's horizontal ($X$) and vertical ($Y$) profiles at various $z$ locations. Good agreement between simulations and measurements validates the beam model implemented in Geant4 simulations.

\subsection{Event Topology}

\begin{figure*}[htbp]
\centering
\includegraphics[width=.99\textwidth]{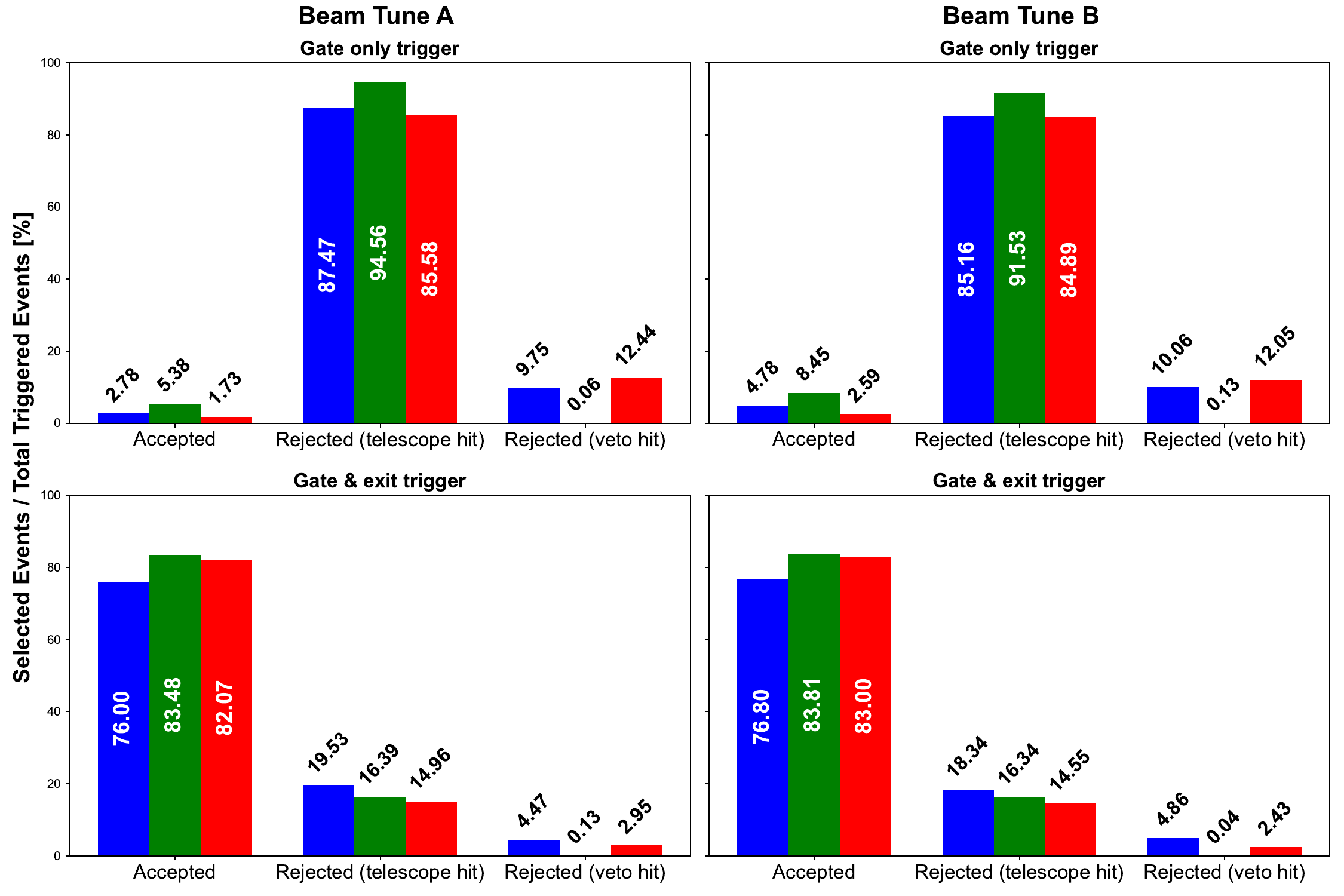}
\caption{Event rate comparisons between experimental data and Geant4 simulation results under different beam tuning (A and B) and trigger conditions (Gate only and Gate \& Exit). The plots quantify the proportions of accepted and rejected events, demonstrating good consistency between measured data and simulations.} 
\label{fig:evRates}
\end{figure*}

Truth-level simulations (simulations excluding optical photon transport) were conducted to investigate muon-detector interactions at the energy deposition level. At this level, the relative event rates for accepted events (muons traversing the gate but not the telescope or veto detector) and rejected events (hits from the gate and telescope detectors, or hits from the gate and veto detectors) reasonably matched experimental trends (Fig.~\ref{fig:evRates}). For example, for beam-tune A and the gate-and-exit trigger sample, the fraction of accepted muons was 76.0\% experimentally compared to 83.48\% in the truth-level simulation. To investigate whether this minor discrepancy between the simulation and measurement arises from the absence of optical processes, simulations incorporating optical photon production and related processes were conducted.

\begin{figure}
    \centering
    \includegraphics[width=.99\textwidth]{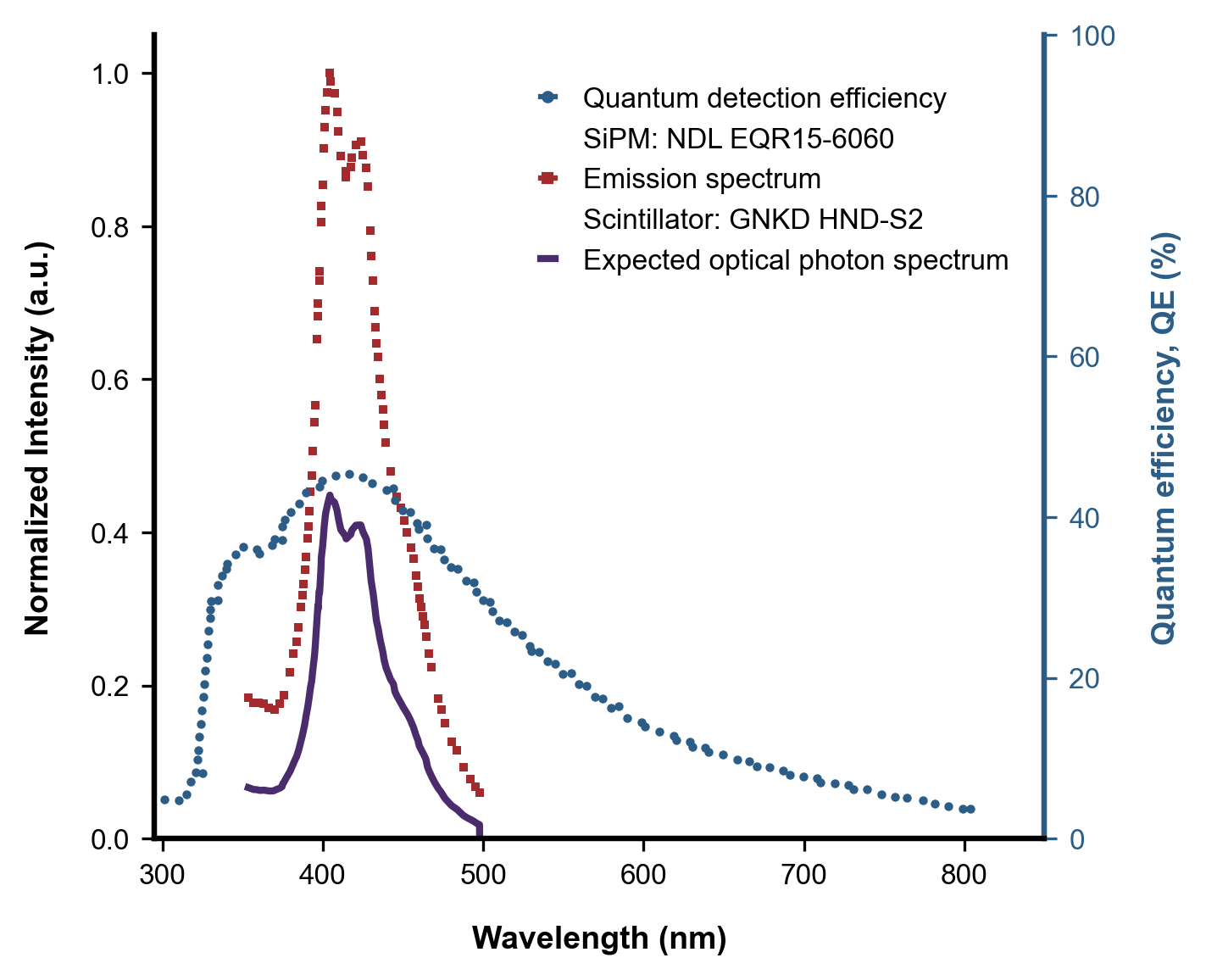}
    \caption{Spectral distribution of photons produced by the GNKD HND-S2 plastic scintillator and the corresponding photon detection efficiency (PDE) of the NDL SiPMs used in the telescope detector. The figure also shows the expected wavelength spectrum of detected photons.}
    \label{fig:wavelengthdistribution}
\end{figure}

Managing optical processes within Geant4 simulations is quite complex, as the generation, transportation, and detection of optical photons rely on various material, surface, and photo-sensor properties. For the plastic scintillators and SiPMs, we have incorporated their properties from the datasheets, and a summary of key parameters is provided below:
\begin{itemize}
    \item GNKD HND-S2 scintillators: Scintillation yield = 8,000\,photons/MeV, attenuation length = 1.2\,m, and emission spectrum peaking at 425\,nm
    \item NDL SiPMs: Photon detection efficiency (PDE) = 40\% at 420\,nm
\end{itemize}
The expected wavelength spectrum of SiPM detected photons is shown in Fig.~\ref{fig:wavelengthdistribution}.

After implementing the plastic scintillator and SiPM properties, our investigation and optimization efforts concentrated on parameters identified in a prior study focused on recognizing critical surface parameters influencing optical photon transport in Geant4~\cite{Nilsson:2015}. In conclusion, material and surface properties constitute the primary parameters, with the surface parameter \textit{finish} identified as the most significant one. A summary of the selected parameters is shown below:
\begin{itemize}
    \item Surface type: dielectric-dielectric (between plastic scintillators), dielectric-metal (between plastic scintillators and SiPMs).
    \item Surface finish: polished air (plastic scintillators), polished (SiPMs)
    \item Surface parameters (plastic scintillators): REFLECTIVITY: 0.95; TRANSMITTANCE: 0.1
    \item Surface parameters (SiPMs): REFLECTIVITY: 0. (all SiPMs); EFFICIENCY: scaled to 0.7 of max PDE (top and right SiPMs); scaled to 0.8 of max PDE (bottom and left SiPMs).
\end{itemize}
All the parameters mentioned above were fine-tuned to match the experimentally measured number of photo-electrons shown in Fig.~\ref{fig:photon_distribution}.

\begin{figure*}[htbp]
\centering
\includegraphics[width=.99\textwidth]{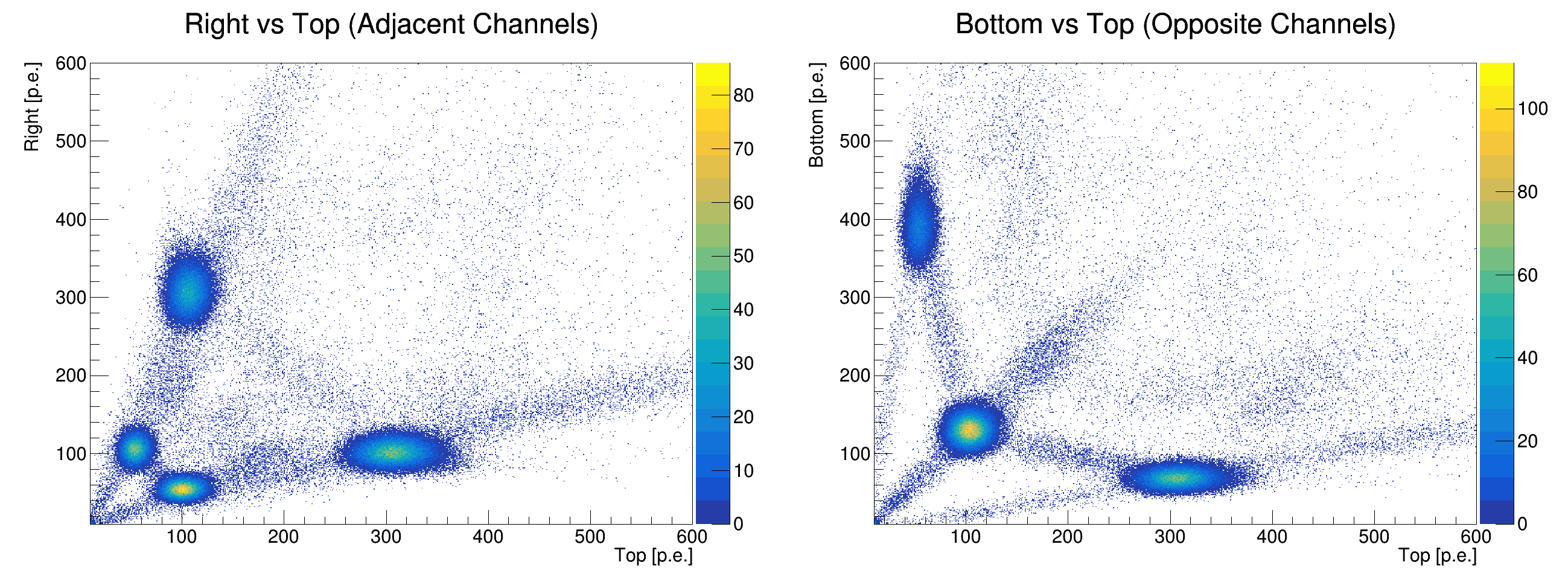}
\caption{Simulated optical photon distribution in the telescope detector obtained from Geant4 simulations, illustrating photo-electron count correlations between adjacent and opposite scintillators. These results show good agreement with the experimental photon distributions in Fig.~\ref{Experimental distribution for CH3 versus CH9} and Fig.~\ref{Experimental distribution for CH3 versus CH7}, validating the optical photon modeling.} 
\label{fig:nPhotDist}
\end{figure*}

Optical cross-talk between scintillators was automatically reproduced by setting the surface parameters aforementioned. A 250\,ns time cut, corresponding to the WaveDAQ data acquisition window, was applied to the simulated photon data to isolate mainly the muon signals. Without this time cut, contamination from decay positrons (with a continuous energy spectrum) would obscure the four blobs (from the 4\,MeV muons) in the photon count correlation plot. Figure~\ref{fig:nPhotDist} illustrates the simulated photon distribution, which reproduces the experimental correlations shown in Fig.~\ref{fig:photon_distribution} (c–d).

Notably, optical-level simulations improved agreement with experimental trigger rates. For gate-exit coincidences, the measured rate of 76\% matched simulations at 72\% (Fig.~\ref{fig:evRates}), while truth-level simulations overestimated this by 8\%. Residual discrepancies likely stem from simplifications in SiPM PDE modeling and minor misalignments in the simulated detector geometry. Finally, the beam focuses (tune A and tune B) have minimal effect on the event rates.

\section{Conclusion}\label{sec:summary}

A prototype telescope detector for the muEDM experiment was successfully constructed using four GNKD HND-S2 scintillator bars and NDL EQR15-6060 SiPMs, and it was tested at the PSI $\pi$E1 beamline with a 27.5 MeV/c muon beam. The event rates were measured under three distinct trigger models, revealing that approximately 2–5\% of the beam muons pass through the gate detector without interacting with the telescope. For muons on the correct trajectory (i.e., passing through both the gate and exit detectors), the triggering efficiency was measured at 75\%.

Double signals were observed, corresponding to muons decaying into positrons within one of the scintillators. These events were confirmed by the characteristic time distribution between the two signals. Photon collection by the SiPMs was consistent with expectations, with more than 300 photons detected by the SiPM attached to the hit scintillator and 50–120 photons recorded by other SiPMs, sufficient to produce the anti-coincidence signal.

The results were cross-validated using Geant4-based simulations, which showed good agreement with the measured event rates at the optical photon level. This agreement demonstrates a robust understanding of the prototype detector's response, providing a solid foundation for further development and commissioning of the muEDM experimental setup.

\section*{Declarations}
\subsection*{Conflict of interest}
On behalf of all authors, the corresponding author states that there is no conflict of interest.

\bmhead{Acknowledgments}
The project is supported by the Science Foundation of China under Grant 12050410233 and the China Scholarship Council No. 202206230093. Additionally, this work is partially funded by the Swiss National Science Fund through grants 204118 and 220487 and receives financial support from the Swiss State Secretariat for Education, Research, and Innovation (SERI) under grant number MB22.00040. We wish to express our sincere gratitude to A. Soter and D. G\"oldi for their efforts in ensuring the SiMon detector was adequately prepared prior to our designated beam time. Furthermore, we acknowledge the significant assistance provided by A. Knecht and A. Antognini before and during our test beam times on the $\pi$E1 beam line.

\bibliography{sn-bibliography}% common bib file

\end{document}